\documentclass[aps,pra,preprint,groupedaddress]{revtex4-2}

\usepackage{xcolor}					

\newcommand{\white}[1]{\textcolor{white}{#1}}
\usepackage{placeins}
\usepackage[utf8]{inputenc}
\usepackage{gensymb}
\usepackage{amsmath,amssymb}
\usepackage{placeins}
\usepackage{multirow}
\usepackage{booktabs}

\usepackage{float}
\usepackage{setspace}

\usepackage[english]{babel}
\makeatletter\AtBeginDocument{\let\@elt\relax}\makeatother

\usepackage{overpic}
\usepackage{xfrac}

\begin{document}

\title{Switch of tonal noise generation mechanisms \\ in airfoil transitional flows}

\author{Tulio R. Ricciardi}
\affiliation{University of Campinas, Campinas, SP, Brazil}

\author{William R. Wolf}
\affiliation{University of Campinas, Campinas, SP, Brazil}


	\begin{abstract}
	Large eddy simulations are performed to study tonal noise generation by a NACA0012 airfoil at an angle of attack ${\alpha = 3}$ deg. and freestream Mach number of ${M_{\infty} = 0.3}$. Different Reynolds numbers are analyzed spanning ${0.5 \times 10^5 \le Re \le 4 \times 10^5}$. Results show that the flow patterns responsible for noise generation appear from different laminar separation bubbles, including one observed over the airfoil suction side and another near the trailing edge, on the pressure side. For lower Reynolds numbers, intermittent vortex dynamics on the suction side results in either coherent structures or turbulent packets advected towards the trailing edge. Such flow dynamics also affects the separation bubbles on the pressure side, which become intermittent. Despite the  irregular occurrence of laminar-turbulent transition, the noise spectrum depicts a main tone with multiple equidistant secondary tones. Increasing the Reynolds number leads to a permanent turbulent regime on the suction side that reduces the coherence level causing only small scale turbulent eddies to be observed. Furthermore, the laminar separation bubble on the suction side almost vanishes while that on the pressure side becomes more pronounced and permanent. As a consequence, the dominant noise generation mechanism becomes the vortex shedding along the wake.
\end{abstract}

\maketitle

\section{Introduction}

	In the early experiments by \citet{Paterson1973}, noise generation by NACA airfoils was investigated for several angles of attack over a Reynolds number range between $10^5<Re<10^6$. Results showed the existence of multiple discrete tones in the acoustic spectra, which were often characterized by a main tonal peak and multiple equidistant secondary tones superposed on top of a broadband hump. This pattern was attributed to trailing-edge acoustic scattering from spanwise coherent instabilities shed from a laminar separation bubble (LSB) on the airfoil pressure side. For the flows investigated, it was argued that the acoustic waves would propagate upstream, exciting new hydrodynamic disturbances that convect downstream, sustaining a self-excited acoustic feedback loop \cite{Tam1974,Arbey1983,Lowson1994}. This flow dynamics has been corroborated by means of linear stability analyses that provide further insights on the sensitivity and response to acoustic perturbations at the frequencies of the tonal noise components \cite{Nash1999, Desquesnes2007, Jones2011, FosasdePando2014, Wu2021, ricciardi2022JFM}.

	The first numerical studies of airfoil secondary tones and acoustic feedback loop mechanisms only considered two-dimensional simulations at moderate Reynolds numbers $0.5\times10^{5} < Re < 2\times10^{5}$ \cite{Desquesnes2007,Jones2011,FosasdePando2014,Golubev2014}. This approach was justified based on the high spanwise flow coherence observed in early experiments, and considering that the important mechanisms for tonal noise generation arise from two-dimensional flow instabilities. These simulations provided valuable insights on the noise sources but could not account for intrinsic three-dimensional (3D) effects, such as vortex stretching and transition to turbulence. To overcome this limitation, 3D simulations have been recently employed to investigate such effects in the context of airfoil secondary tones and the feedback loop mechanisms \cite{Sanjose2018,gelot_kim_2020,Golubev2021,Wu2021,ricciardi2022JFM}.

	Early experiments were usually conducted at low angles of incidence and high Reynolds numbers, in the range of $3\times10^5 < Re < 1.5\times10^6$ \cite{Nash1999,Plogmann2013}. At such flow conditions, the boundary layers on the suction side were typically turbulent while, on the bottom side, they remained in a laminar regime due to the favorable pressure gradient. However, near the trailing edge, the pressure gradient may also become adverse, leading to a local flow detachment on the pressure side depending on the airfoil angle of attack. In these cases, flow instabilities arising on the airfoil pressure side, close to the trailing edge, lead to shedding of coherent structures and tonal noise generation due to acoustic scattering. It has been shown that the tonal component vanishes if tripping devices are applied to force turbulent transition on the pressure side boundary layers \cite{Wolf2012, Plogmann2013,Probsting2015_bubble,Probsting2015_regimes,Yakhina2020}.

	For lower Reynolds numbers, $0.4\times10^5 < Re < 3\times10^5$ \cite{Probsting2015_regimes}, the suction side boundary layer may also detach due to the adverse pressure gradient, leading to LSBs on both sides of the airfoil. In these cases, the coherence level of flow structures reaching the trailing edge dictates if one side is solely responsible for the tonal noise generation or if flow structures from both sides interact, leading to a more complex noise generation mechanism \cite{Desquesnes2007, Probsting2014}. This intricate aeroacoustic phenomenon is sensitive to parameters such as the angle of attack and the Reynolds number \cite{Probsting2014, Probsting2015_regimes,Golubev2021}, besides the airfoil geometry \cite{Yakhina2020} and compressibility effects \cite{Ricciardi_Scitech2021}. When combined, these parameters will dictate the physical mechanisms responsible for airfoil tonal noise generation including the detachment and reattachment dynamics of separation bubbles, and the subsequent shedding of coherent structures. For instance, it has been reported that vortex merging \cite{yarusevich2019_merging, ricciardi2022JFM}, bursting \cite{jones2008, jones2010, yarusevich2016_coherent, yarusevych2018_transition, Sanjose2018, jaiswal2022}, and intermittency \cite{Desquesnes2007, Probsting2014, Sanjose2018, ricciardi2022JFM} take place downstream of suction side separation bubbles. Thus, it is clear that the investigation of the flow dynamics driven by the separation bubbles is important to understand the noise sources.

	Since different physical processes dominate depending on angle of attack and Reynolds number, \citet{Probsting2015_regimes} performed experiments for a NACA0012 airfoil at effective angles of attack $0 \le \alpha \le 6.3$ deg, for Reynolds numbers $0.3\times10^{5} \le Re \le 2.3\times10^{5}$. The previous authors demonstrated that, as the Reynolds number increases, the noise generation mechanisms switch sides on the airfoil. It was observed that lower Reynolds numbers lead to more pronounced suction side events while higher Reynolds numbers are dominated by the pressure side dynamics. Moreover, it was also shown that the magnitude of secondary peaks reduce as the suction side boundary layer becomes turbulent. Recent experiments applied time-frequency analyses to the acoustic signals and identified different patterns of noise emission for transitional airfoils, which depicted single and multiple tones depending on the specific flow conditions. For some cases, tonal noise was not observed, confirming that airfoil noise generation is indeed very dependent on the flow parameters \cite{Probsting2015_bubble,Padois2016,Yakhina2020,Golubev2021}.

	In order to study the Reynolds number dependence on airfoil trailing-edge noise, we employ wall-resolved large eddy simulations (LES) with particular attention to the behavior of secondary tones and their sources. The main contribution of the present study relates to the characterization of hydrodynamic changes that take place on the suction and pressure side boundary layers at transitional flows. In the present simulations, the boundary layers develop naturally, without any tripping devices, in order to understand the role of coherent structures at both sides of the airfoil on the noise generation process. An assessment is performed for a NACA0012 airfoil considering four different chord-based Reynolds numbers, including $Re = 0.5, 1$, $2$ and $4 \times 10^5$. The freestream Mach number is set as $M_\infty = 0.3$ and the angle of attack is fixed at $\alpha = 3$ deg. The latter is justified since, at low angles of attack, different patterns of noise radiation are observed depending solely on the Reynolds number \cite{Probsting2015_regimes,Yakhina2020}. It is also important to mention that the present simulations are performed at higher Mach numbers than those typically analyzed in experimental work. The influence of compressibility effects on tonal noise can be found in Ref. \cite{Ricciardi_Scitech2021}.

	Results in terms of mean flow contours of streamwise velocity component $u$ show the presence of two laminar separation bubbles, one along the suction side and another near the trailing edge, on the pressure side. Their characteristic length scales depend on the Reynolds number and play a major role in the onset of coherent structures. The dynamics of spanwise-correlated flow structures on each side of the airfoil and their role in the noise generation are investigated using proper orthogonal decomposition and spectral analysis. The latter is conducted using Fourier transforms for spanwise-filtered and fully 3D disturbances. Intermittency of flow structures is also characterized combining flow visualization of spanwise-filtered snapshots, spanwise covariances and continuous wavelet spectrograms. These techniques allow the characterization of the main frequencies of tonal noise for each flow setup including their relation to the coherent structures advecting at the trailing edge. Finally, the dynamics of the pressure side bubble is also investigated including its role on vortex shedding.

    \clearpage
\section{Numerical Methodology}

	\subsection{Large eddy simulations}

		Large eddy simulations are performed to solve the compressible Navier-Stokes equations in general curvilinear coordinates. The spatial discretization of the governing equations employs a sixth-order accurate compact scheme for derivatives and interpolations on a staggered O-grid \cite{Nagarajan2003}. Away from the boundary layer, a sixth-order compact filter \cite{Lele1992} is applied to control high-wavenumber numerical instabilities arising from grid stretching and interpolation between staggered grids. Since the transfer function associated with such filters provides an approximation to sub-grid scale models \cite{Mathew_etal_2003}, no explicit modeling is employed here. The time integration near the wall is performed by the implicit second-order scheme of Beam and Warming \cite{Beam1978} while an explicit third-order Runge-Kutta scheme is used in flow regions away from the solid boundary. The communication across the different time marching schemes is performed at overlapping grid points. Periodic boundary conditions are used in the spanwise direction while sponge plus characteristic boundary conditions are applied in the far-field locations to avoid wave reflections. No-slip adiabatic wall boundary conditions are enforced along the airfoil surface.

		Length scales, velocity components, density and pressure are non-dimensionalized as $\boldsymbol{x} = \boldsymbol{x}^*/L^*$, $\boldsymbol{u}=\boldsymbol{u}^*/a_{\infty}^*$, $\rho=\rho^*/{\rho}_{\infty}^*$ and $p = p^*/{\rho}_{\infty}^*{a_{\infty}^*}^2$, respectively. Here, $L^*$ is the airfoil chord, $a_{\infty}^*$ is the freestream speed of sound and $\rho_{\infty}^*$ is the freestream density, and the quantities with superscript $^{*}$ are given in dimensional units. Herein, time and frequency are presented in terms of convective time units (CTUs) and Strouhal number. Therefore, they are non-dimensionalized by freestream velocity as $t=t^* \, U_{\infty}^* / L^*$ and $St=f^* \, L^* / U_{\infty}^*$, respectively. The present numerical tool has been validated for several simulations of compressible flows around turbine blades and airfoils \cite{Bhaskaran2010, Wolf2012, Wolf:DU96, Wolf2013, Brener2019, Miotto2022}.

\begin{figure}[H]
   			\centering
   			\begin{overpic}[width=0.99\textwidth]{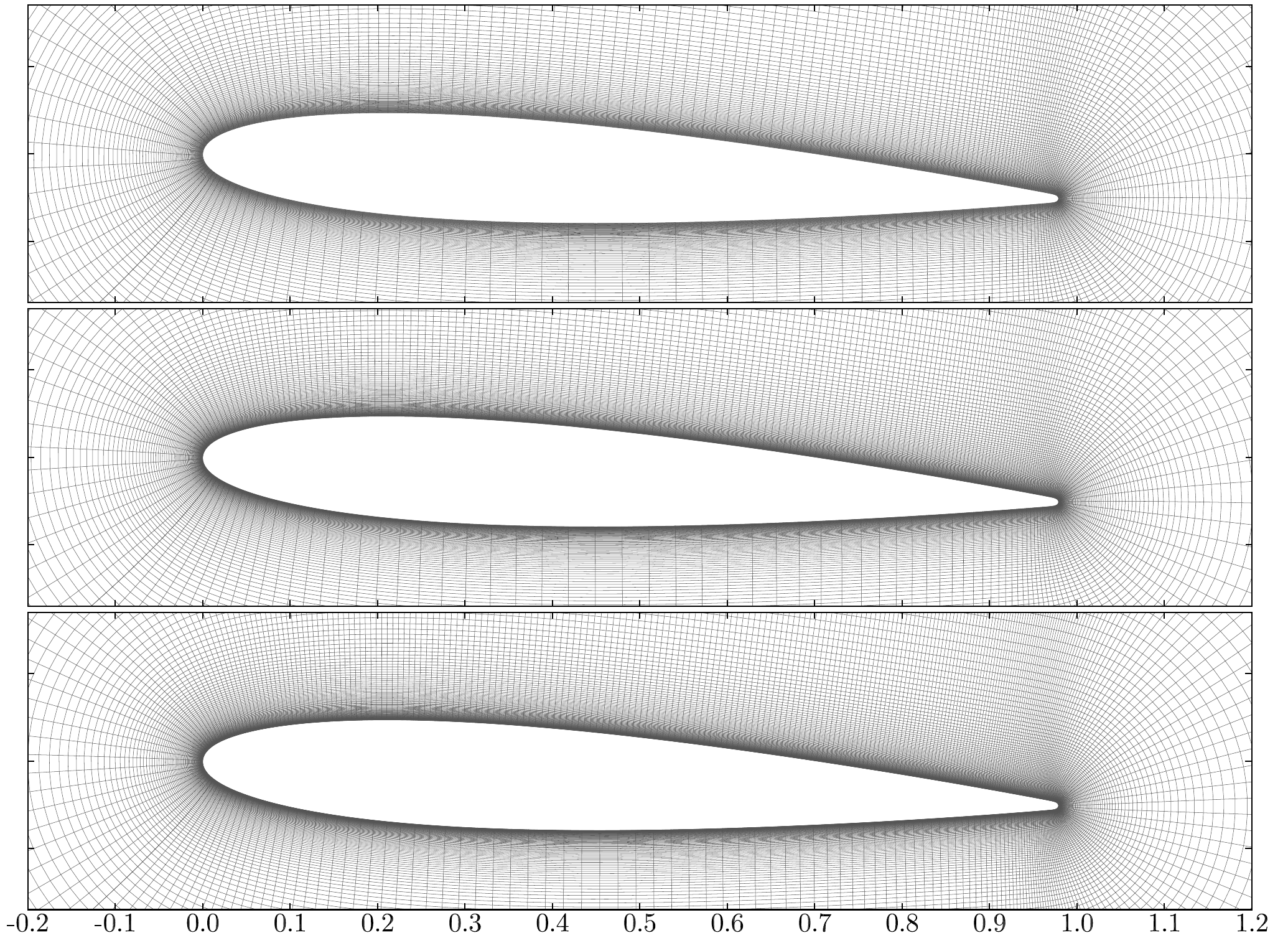}
   				\put(2.3,50.8){\colorbox{white}{\parbox{0.025\linewidth}{(a)}}}
   				\put(2.3,27.2){\colorbox{white}{\parbox{0.025\linewidth}{(b)}}}
   				\put(2.3,3.7){\colorbox{white}{\parbox{0.025\linewidth}{(c)}}}
   			\end{overpic}
   			\caption{Computational grids near the airfoil for (a) $Re = 0.5$ and $1 \times 10^5$, (b) $Re = 2\times 10^5$ and (c) $Re = 4 \times 10^5$ are shown skipping every three points in both directions.}
   			\label{fig:grid}
   		\end{figure}

		For a smooth O-grid generation, the original NACA0012 airfoil with a blunt trailing edge is truncated at 98\% of the reference chord $L^{*}$ and it is rounded with a curvature radius of $r/L^{*} = 0.4\%$. Thus, hereafter, the non-dimensional chord of the modified airfoil is given by $c = 0.98$. The leading edge is placed at $(x, y) = (0, 0)$ such that the airfoil is pivoted about this point. Three different computational O-grids are employed for the current LES and they are shown in gray lines for every three points in Fig. \ref{fig:grid}. The first mesh is presented in Fig. \ref{fig:grid}(a) and it is used for the low Reynolds number configurations, $Re = 0.5$ and $1 \times 10^5$. The mesh employed in the intermediate Reynolds number, $Re = 2 \times 10^5$, is shown in Fig. \ref{fig:grid}(b). For the highest Reynolds number, $Re = 4 \times 10^5$, the grid is shown in Fig. \ref{fig:grid}(c).

\begin{figure}[H]
   			\centering
   			\begin{overpic}[width=0.99\textwidth]{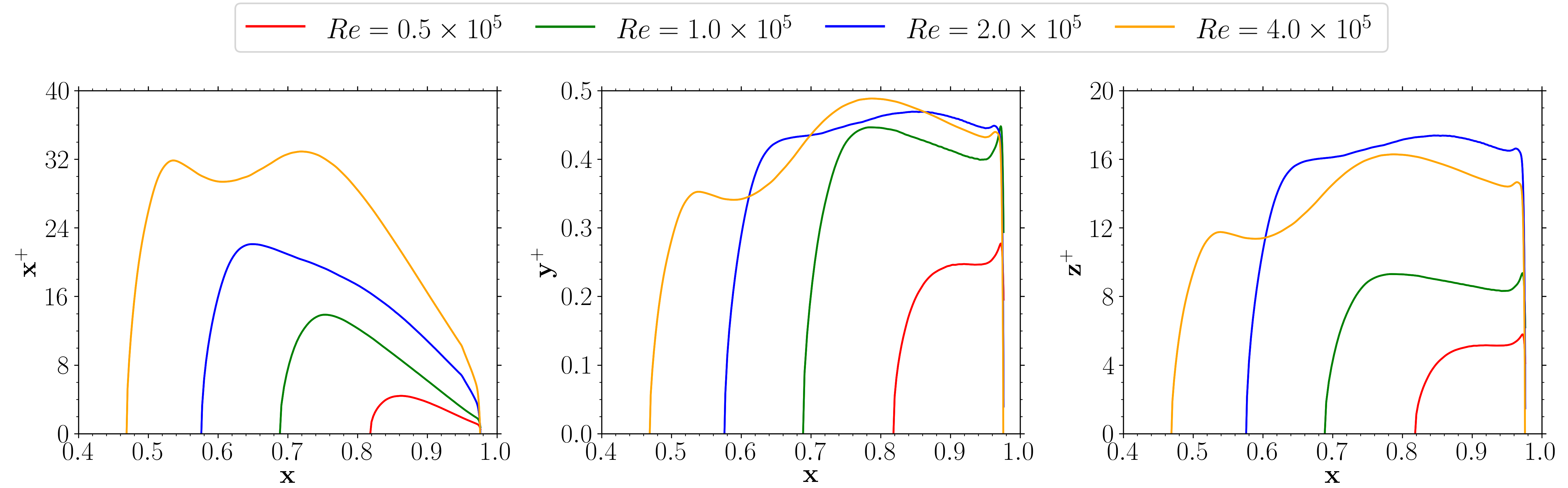}
   				\put(5.5,22.7){(a)}
   				\put(39,22.7){(b)}
   				\put(72.5,22.7){(c)}
   			\end{overpic}
   			\caption{Grid resolution along the turbulent boundary layer region on the suction side in terms of (a) streamwise, (b) wall-normal and (c) spanwise wall units.}
   			\label{fig:mesh_refin}
   		\end{figure}

		As discussed in earlier numerical studies \cite{Desquesnes2007,Jones2011,FosasdePando2014}, the important mechanisms for tonal noise generation arise from two-dimensional flow instabilities in the laminar flow region. Motivated by this observation, a grid refinement study was conducted in terms of mean and fluctuation properties for 2D simulations considering $Re = 0.5$ and $1\times10^5$ \cite{Ricciardi2019_tones,Ricciardi_Scitech2021}. From these previous references, we conclude that the present mesh setups have an adequate resolution to compute the laminar instabilities. Due to the transition to turbulence downstream of the LSB, the assessment of grid resolution along the turbulent boundary layer is performed in terms of wall units. Results are presented in Fig. \ref{fig:mesh_refin} for streamwise, wall-normal and spanwise wall units, $x^+$, $y^+$ and $z^+$, respectively. The current values are in accordance with the recommended values for wall-resolved LES \cite{Georgiadis2010}. We also employ spanwise domains which are wider than those from recent studies \cite{Sanjose2018,gelot_kim_2020,Golubev2021}.

		The 3D simulations start from 2D flows superposed with initial random noise. The flow transients are discarded after they achieve a statistically stationary condition. After that, the simulations were conducted for long time periods in order to capture different intermittent flow patterns and their impact on the instantaneous noise generation \cite{Padois2016,Sanjose2018,ricciardi2022JFM}.

		Additional details on the computational grids and simulation parameters are presented in table \ref{table1}. It presents the number of grid points along the streamwise, $n_{\xi}$, wall-normal, $n_{\eta}$, and spanwise directions, $n_{z}$. Further details are also described in terms of grid resolution at the wall, $\Delta \eta_w$, total spanwise $L_z$ and radial size of the domain, as well as the time step per iteration, $\Delta t$, and the total time of the simulation in convective time units.

		\begin{table}[H]
        	\centering
        	\caption{Description of flow configurations investigated.}
        	\label{table1}
        	\begin{tabular}{|c|c|c|c|c|}
        		\hline
        		$\boldsymbol{Re}$                           & $0.5 \times 10^5$   & $1 \times 10^5$ & $2 \times 10^5$  & $4 \times 10^5$ \\ \hline
        		$\boldsymbol{M_\infty}$                     & \multicolumn{4}{c|}{0.3}                                                   \\ \hline
        		$\boldsymbol{\alpha}$                       & \multicolumn{4}{c|}{3 deg.}                                                \\ \hline
        		$\boldsymbol{n_\xi}$                        & \multicolumn{2}{c|}{660}              & 792              & 1080            \\ \hline
        		$\boldsymbol{n_\eta}$                       & \multicolumn{2}{c|}{600}              & 480              & 520             \\ \hline
        		$\boldsymbol{n_z}$                          & \multicolumn{2}{c|}{192}              & 216              & 240             \\ \hline
        		$\boldsymbol{\Delta \eta_w \times 10^{5}}$  & \multicolumn{2}{c|}{10}               & 5                & 2.5             \\ \hline
        		$\mathbf{L_z}$                              & \multicolumn{3}{c|}{0.4c}                                & 0.2c            \\ \hline
        		\textbf{Domain size}                        & \multicolumn{2}{c|}{36c}              & \multicolumn{2}{c|}{24c}           \\ \hline
        		$\boldsymbol{\Delta t \times 10^{4}}$       & \multicolumn{2}{c|}{1.5}              & 1.0              & 0.5             \\ \hline
        		\textbf{CTU}                                & \multicolumn{3}{c|}{75}                                  & 30              \\ \hline
        	\end{tabular}
        \end{table}

    \clearpage
	\subsection{Analysis of coherent structures}
	  	\label{sec:pod_form}

		The identification of organized patterns is an important task in fluid mechanics since coherent flow structures can be responsible for a large portion of the flow kinetic energy. Furthermore, spanwise coherent structures may lead to an efficient sound generation due to trailing edge scattering \cite{FWHall1970, sanoprf2019}. Coherent structures are defined in terms of fluctuation quantities, such that the Reynolds decomposition is employed to split the instantaneous flow field ${q}(\mathbf{x},t)$ into an averaged time-independent component ${\bar{q}}(\mathbf{x})$ and its fluctuation field ${q}'(\mathbf{x},t)$. Due to the spatial and temporal discretizations in the simulations, the flow field $q$, the spatial coordinates $\mathbf{x}$ and the time $t$ are treated as discrete variables in the following equations.

		In this work, the organized flow patterns are identified using two different techniques including the two-point, one-time autocovariance along the spanwise direction, defined as
		\begin{equation}
		R(x,y,\Delta z,t) = \left\langle q'(x,y,z,t) \, q'(x,y,z + \Delta z,t) \right\rangle_z  \mbox{ ,}
		\label{eq:correlation}
		\end{equation}
		and the proper orthogonal decomposition (POD) \cite{Lumley1967,Sirovich1990}. The latter is a data-driven method that seeks the best $L_2$-norm representation of the fluctuation flow field. It requires the calculation of the covariance matrix $\mathbf{C}$, defined as
		\begin{equation}
		\mathbf{C} = \mathbf{Q'\,}^{T} \mathbf{W} \ \mathbf{Q'} \mbox{ ,}
		\end{equation}
		where the matrix $\mathbf{Q}'$ contains the fluctuation field ${q}'(\mathbf{x},t)$ and its transpose $\mathbf{Q}'^{T}$, and $\mathbf{W}$ contains the integration weights based on the spatial discretization.
		The matrix $\mathbf{C}$ is then decomposed using the singular value decomposition (SVD) and yields orthogonal empirical functions $\boldsymbol{\phi}_{i}(\mathbf{x})$ and $\mathbf{a}_{i}(t)$, respectively the spatial and temporal modes, with an amplitude scaling factor given by the singular values $\boldsymbol{\sigma}_i$.
		The modes can be recombined linearly and without loss of information according to
		\begin{equation}
		\mathbf{Q}'(\mathbf{x},t) = \sum_{i=1}^{M} \mathbf{a}_{i}(t) \boldsymbol{\sigma}_i \boldsymbol{\phi}_{i}(\mathbf{x}) \mbox{.}
		\end{equation}
		In this decomposition, the maximum number $M$ of modes is equal to the rank of $\mathbf{Q}'$, i.e., the smallest number of samples in space or time.
		In order to perform a comparison between different modes in the results section, the singular values are incorporated into the temporal dynamics.

    \clearpage
\section{Results and Discussion}

	The physical mechanisms responsible for airfoil noise generation on transitional flows are discussed in this section. For this task, flow visualization and signal processing techniques are employed to investigate the role of Reynolds number on the airfoil suction and pressure side events.

	\subsection{Flow dynamics}

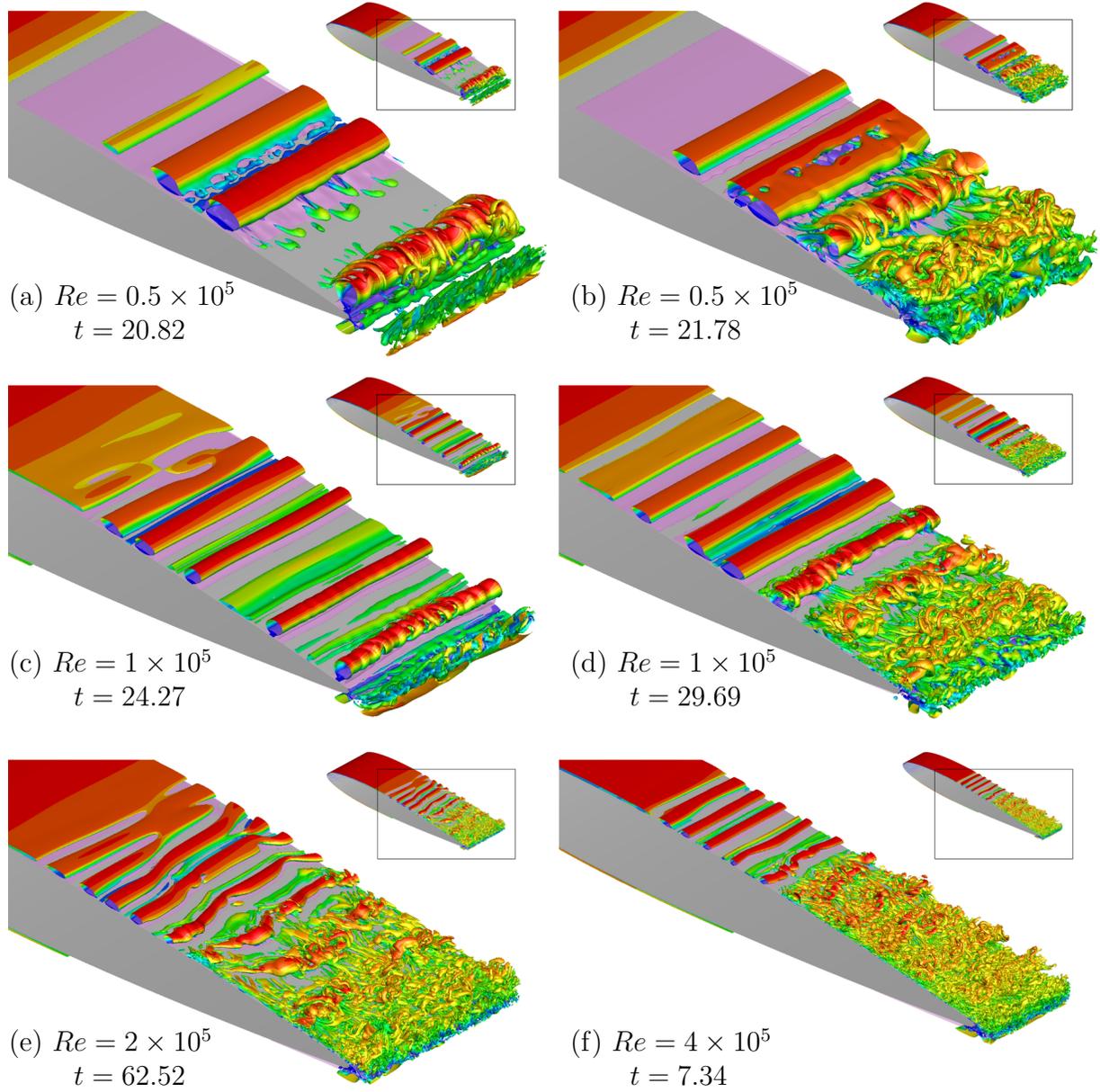
\begin{figure}
			\centering
			\begin{overpic}[width=0.99\textwidth]{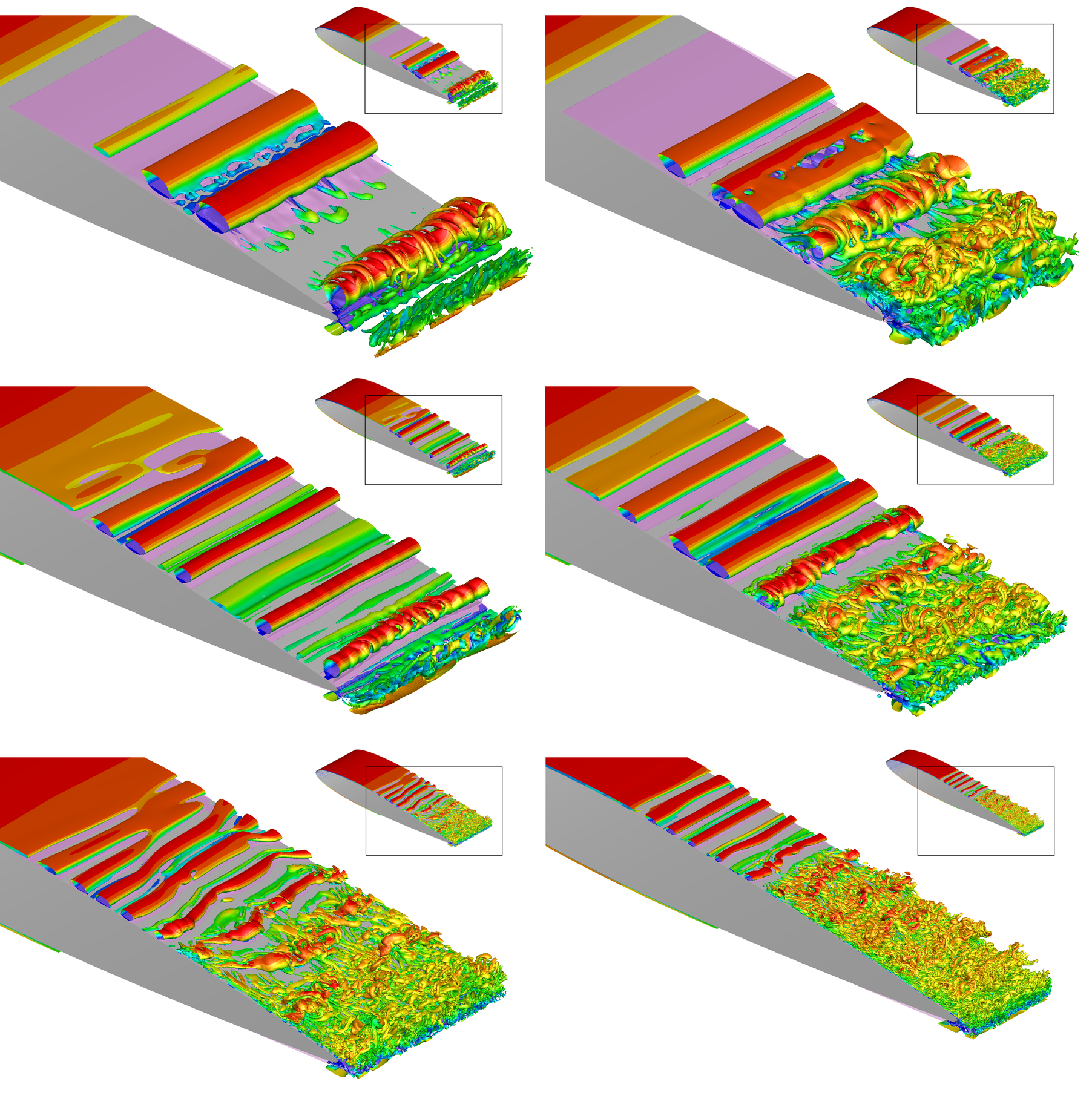}
				\put(0,72.5){(a) $Re = 0.5\times10^5$}
				\put(5.8,69.5){$t = 20.82$}
				\put(50.0,72.5){(b) $Re = 0.5\times10^5$}
				\put(55.3,69.5){$t = 21.78$}
				\put(0,39.8){(c) $Re = 1 \times 10^5$}
				\put(5.8,36.8){$t = 24.27$}
				\put(50.0,39.8){(d) $Re = 1 \times 10^5$}
				\put(55.3,36.8){$t = 29.69$}
				\put(0.0,6){(e) $Re = 2 \times 10^5$}
				\put(5.8,3){$t = 62.52$}
				\put(50.0,6){(f) $Re = 4 \times 10^5$}
				\put(55.3,3){$t = 7.34$}
			\end{overpic}
			\caption{Iso-surface of Q-criterion colored by streamwise velocity component. The magenta shaded surfaces indicate instantaneous reversed flow regions which highlight the size of the LSB.}
			\label{fig:Qcriterion}
		\end{figure}

		As an initial assessment of the flow dynamics, the visualization of solution snapshots is presented in Fig. \ref{fig:Qcriterion} in terms of isosurfaces of the Q criterion (Q $ = 2.5$) colored by the streamwise velocity component $u$. In the plots, a magenta shaded region indicates instantaneous reversed flow in order to highlight the LSB. These pictures are extracted from movie \#1 in the supplemental material, where the flow dynamics is better visualized.

		For the lowest Reynolds number, $Re = 0.5 \times 10^5$, the overall dynamics is dictated by vortex pairing on the suction side \cite{ricciardi2022JFM}. The flow instabilities that originate near mid-chord are purely 2D but, as they are advected, interaction among vortices may introduce three-dimensionality to the structures that eventually leads to vortex breakdown. This process results in flow intermittency at the trailing edge and it is possible to see in Figs. \ref{fig:Qcriterion}(a) and (b), respectively, a spanwise coherent structure or small-scale turbulent eddies. The coherent structures observed in the present study are similar to those previously reported in both numerical simulations and experiments \cite{Sanjose2018,jaiswal2022}.

		With an increase in the Reynolds number to $1 \times 10^5$, different flow patterns are still observed. For instance, coherent structures at the trailing edge are illustrated in Fig. \ref{fig:Qcriterion}(c) while Fig. \ref{fig:Qcriterion}(d) shows turbulent packets with small-scale eddies. The presence of non-zero spanwise wavenumbers near the mid-chord region, during the onset of flow instabilities, as illustrated in Fig. \ref{fig:Qcriterion}(c), is an important difference from this setup compared to the lowest Reynolds number case. This has a direct influence on the vortex pairing because it is more likely that this process results in breakdown rather than merging. The only coherent structures reaching the trailing edge are those that originated from flow instabilities with zero spanwise wavenumber.

		For higher Reynolds numbers, the flows always transition to turbulence along the airfoil suction side and no laminar-like coherent structures are observed to reach the trailing edge, as illustrated in Figs. \ref{fig:Qcriterion}(e) and (f) for $Re = 2$ and $4 \times 10^5$, respectively. As presented in Fig. \ref{fig:Qcriterion}(e), non-zero wavenumbers can be observed in the vortices for $Re = 2 \times 10^5$. For $Re = 4 \times 10^5$, as presented by movie \#1, the flow instabilities still exhibit non-zero wavenumbers, and always transition to turbulence despite their shorter spanwise length.

    \clearpage
	\subsection{Mean flow statistics}

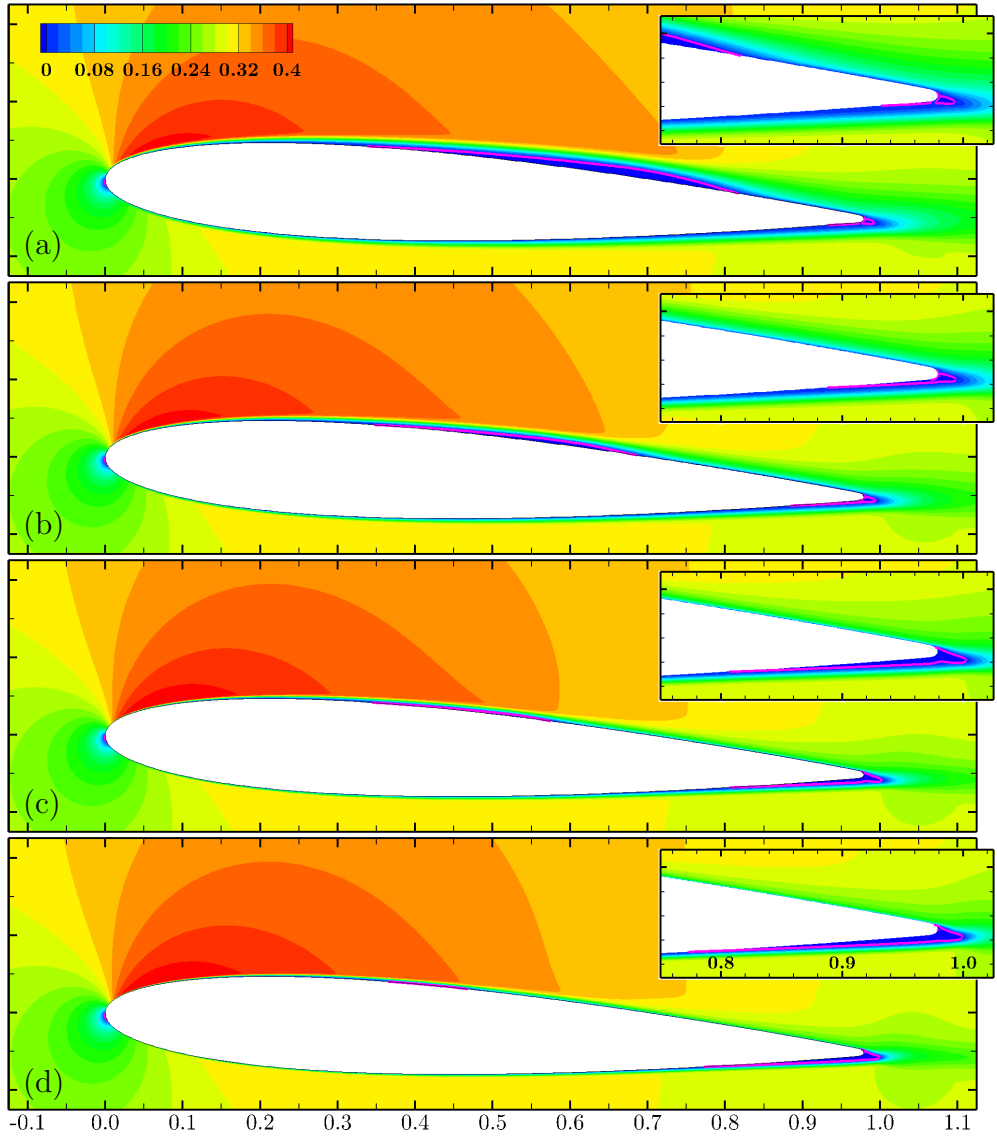
\begin{figure}[H]
			\centering
			\begin{overpic}[width=0.8\textwidth]{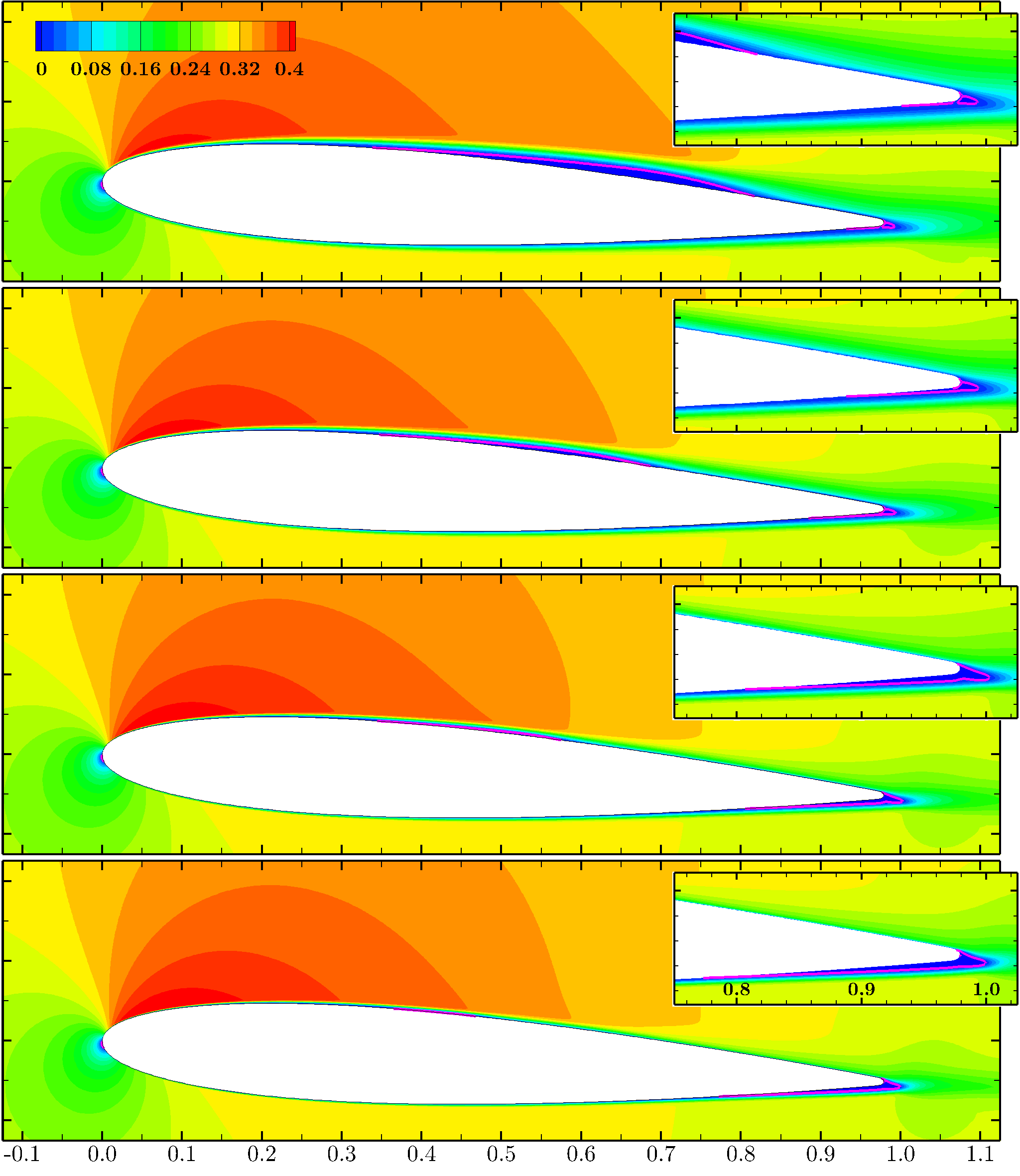}
				\put(1.5,78){(a)}
				\put(1.5,53.5){(b)}
				\put(1.5,29){(c)}
				\put(1.5,4.5){(d)}
			\end{overpic}
			\caption{Mean flow contours in terms of streamwise velocity component ${u}$ for (a) $Re = 0.5\times 10^5$, (b) $Re = 1\times 10^5$, (c) $ Re = 2\times 10^5$ and (d) $Re = 4 \times 10^5$. The magenta lines mark the LSBs based on the averaged reversed flow region in terms of $\bar{u} < 0$.}
			\label{fig:mean_flow}
		\end{figure}

		The mean flow is presented in Fig. \ref{fig:mean_flow} as contours of the streamwise velocity component $u$ normalized by the sound speed. The laminar separation bubbles are observed as dark blue contours bounded by magenta lines, highlighting the reversed flow region. For higher Reynolds numbers, the suction side bubbles are almost imperceptible. This figure shows that, as the Reynolds number increases, the suction side LSB shortens, leading to an earlier flow reattachment. On the other hand, the pressure side LSB becomes more pronounced due to the earlier flow detachment, as can be seen in the magnified views.

\begin{figure}[H]
			\centering
			\begin{overpic}[width=0.99\textwidth]{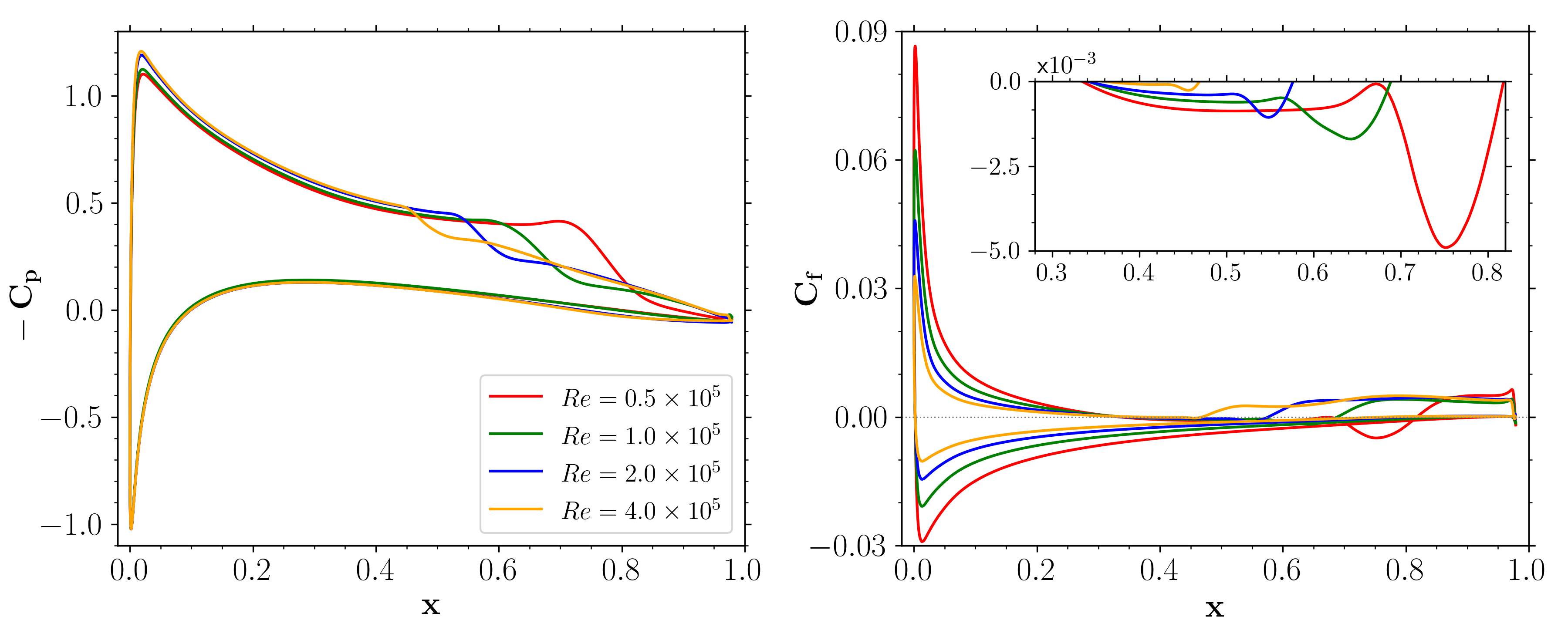}
				\put(0.0,37){(a)}
				\put(49,37){(b)}
			\end{overpic}
			\caption{Pressure and friction coefficients on the airfoil surface.}
			\label{fig:Cp_and_Cf}
		\end{figure}

		Pressure coefficient distributions $\left(Cp = \frac{p - p_\infty}{0.5 \rho_\infty U_\infty^2}\right)$ are presented in Fig. \ref{fig:Cp_and_Cf}(a) for all simulations. Due to the airfoil incidence, the flow has a higher acceleration on the suction side, resulting in large values of $-Cp$ at $x \approx 0.02$. From this point onward, the adverse pressure gradient starts to slow down the flow and eventually leads to a laminar separation bubble on the suction side. The pressure plateaus are more pronounced downstream according to the bubble sizes. The friction coefficient distributions $\left(Cf = \frac{\tau_{w}}{0.5 \rho_\infty U_\infty^2}\right)$ are presented in Fig. \ref{fig:Cp_and_Cf}(b), where $\tau_{w}$ represents the wall shear stress. From this figure, it is possible to see that the flows detach at similar locations, $0.34 \lesssim x \lesssim 0.38$. The flow reattachment, on the other hand, strongly depends on the Reynolds number and occurs at more upstream positions as this parameter increases. This observation is in agreement with recent work \cite{Yakhina2020, Golubev2021}.

\begin{figure}[H]
			\centering
			\begin{overpic}[width=0.75\textwidth]{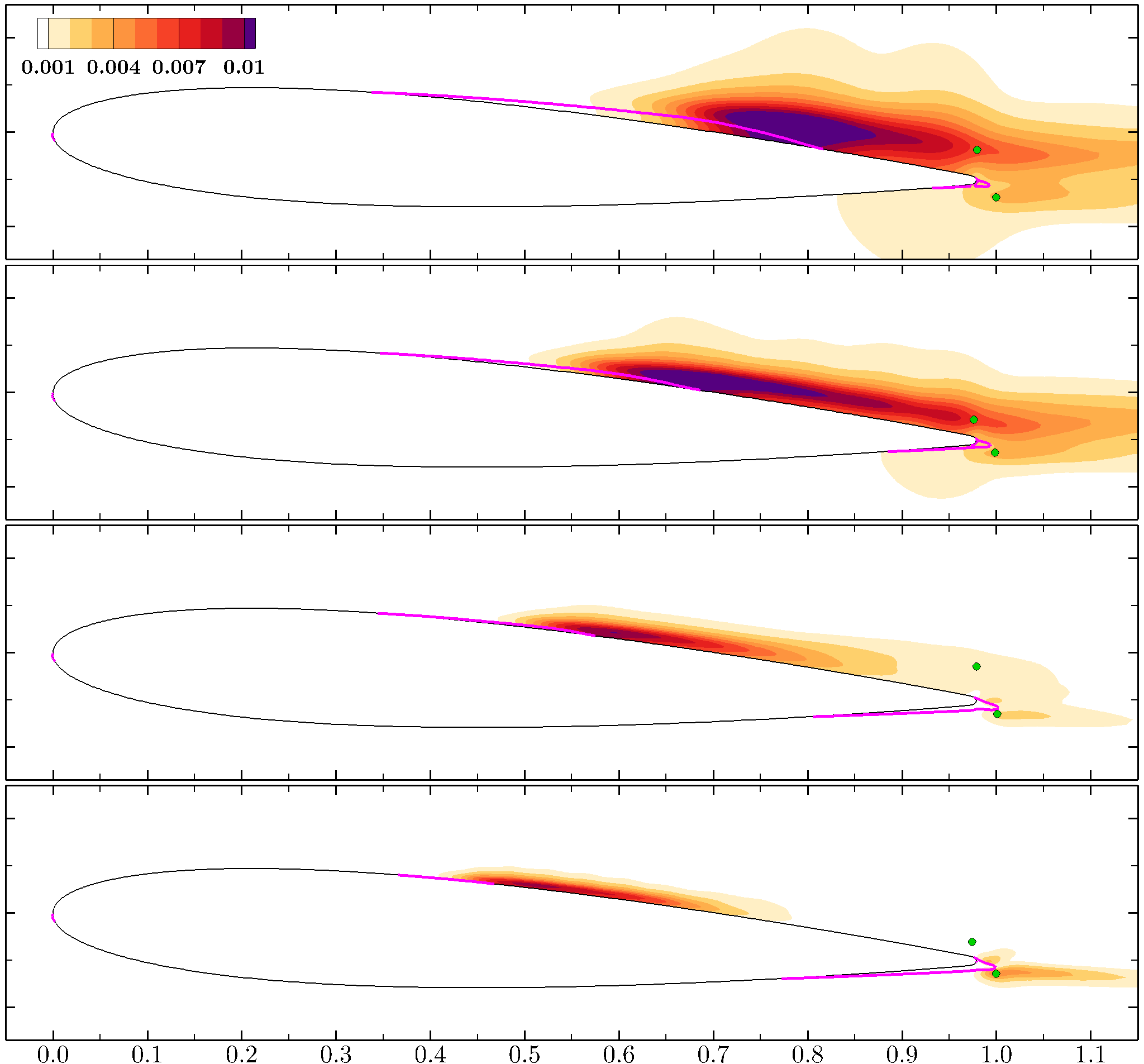}
				\put(2,72.0){(a)}
				\put(2,49.3){(b)}
				\put(2,26.6){(c)}
				\put(2,3.9){(d)}
			\end{overpic}
			\caption{Contours of root-mean square (RMS) values of spanwise-averaged pressure for (a) $Re = 0.5\times 10^5$, (b) $Re = 1\times 10^5$, (c) $Re = 2\times 10^5$ and (d) $Re = 4 \times 10^5$. The magenta lines mark the LSBs based on the averaged reversed flow region in terms of $\bar{u} < 0$. The green symbols are locations used for extraction of temporal signals.}
			\label{fig:Prms_flow}
		\end{figure}
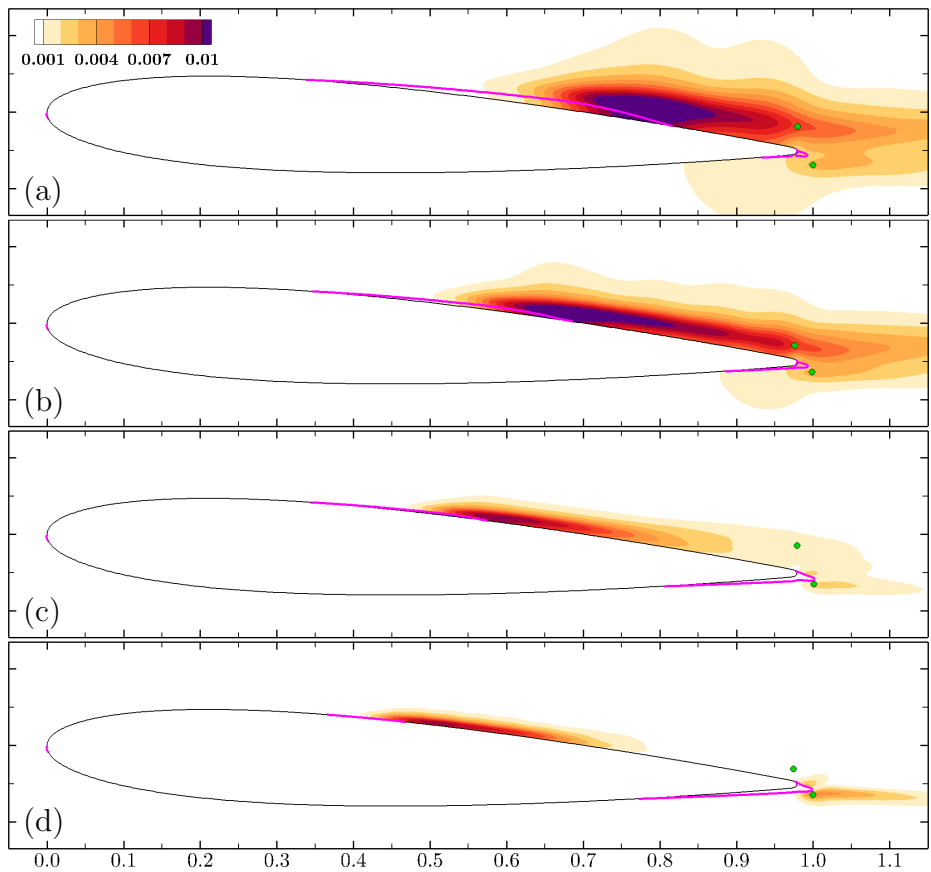

		The root-mean square (RMS) values of the spanwise-averaged pressure field are shown in Fig. \ref{fig:Prms_flow} for all flows investigated. With the spanwise averaging, the contribution from turbulent eddies in the RMS value is reduced since the averaging process acts as a filtering operation on uncorrelated data. Results show similar trends on the suction side for all Reynolds numbers. For instance, the fluctuations are more intense downstream of the suction side LSB, where the flow is characterized by the vortex shedding. However, the lower Reynolds number solutions ($Re = 0.5$ and $1 \times 10^5$) exhibit more intense fluctuations extending along the suction side, including the region in the vicinity of the trailing edge. For the higher Reynolds numbers ($Re = 2$ and $4 \times 10^5$), the RMS levels on the suction side are lower near the trailing edge as a consequence of the reduced spanwise-coherent fluctuations due to turbulence transition. For these cases, the RMS levels near the trailing edge, on the pressure side, are higher than those observed on the suction side, particularly for $Re = 4 \times 10^5$. This behavior indicates that the role of the suction side in the noise generation process is reduced while the pressure side becomes dominant. Finally, the green symbols in this figure mark the reference locations where temporal signals are extracted for spectral analyses, as discussed in the following sections.


    \clearpage
	\subsection{Spectral analysis}
	\label{sec:SpectralAnalysis}

		Spectral analyses of the present flows provide further insights of their dynamics and noise generation mechanisms. For this task, temporal signals of hydrodynamic pressure fluctuations are extracted in the vicinity of the trailing edge, at $x = 0.98 = 1c$ for the suction side, and $x = 1.0$ on the near wake at the bottom side. As indicated by the green symbols in Fig. \ref{fig:Prms_flow}, the $y$-coordinates are selected to coincide with the largest values of $p_{\text{RMS}}$ for each flow configuration. At the acoustic near field, pressure signals are extracted one chord above the airfoil trailing edge at $x = y = 1c$.

		The analysis is performed with Fourier transforms of segments from the entire signal, followed by their averaging in the frequency domain. The segments have an overlap of 66\% and the Hanning window is used to remove edge discontinuities. The frequency resolutions are $\Delta St = 0.05$ for $Re \le 2 \times 10^5$ and $\Delta St = 0.10$ for $Re = 4 \times 10^5$. In the hydrodynamic field, two different approaches are employed in the spectral analysis. The first one computes the Fourier transform of the spanwise-averaged signal (2D data) while the second approach computes a Fourier transform for each one of the $n_z$ grid points along the span, followed by a spanwise averaging of the spectra (3D data). These are given, respectively, by
		\begin{eqnarray}
		\hat{q}(x,y,\omega)_{2D} &= \mathcal{F}\left\lbrace \left\langle q(x,y,t) \right\rangle_z  \right\rbrace \text{ and} \label{eq:fft_1} \\
		\hat{q}(x,y,\omega)_{3D} &= \left\langle \mathcal{F}  \left\lbrace q(x,y,z,t)  \right\rbrace \right\rangle_z \label{eq:fft_2} \mbox{ ,}
		\end{eqnarray}
		where $\mathcal{F}$ denotes the Fourier transform and $\left\langle \,\, \right\rangle_z$ represents the spanwise averaging process. Thus, it is possible to estimate the spectral distribution from the spanwise-coherent structures using Eq. (\ref{eq:fft_1}) relative to that of the fully three-dimensional flow from Eq. (\ref{eq:fft_2}). The former acts as a filter that includes the flow disturbances which are more relevant to the acoustic scattering process as discussed by \citet{FWHall1970}, while the latter also includes effects from streamwise correlated, e.g., boundary layer streaks, and other uncorrelated fluctuations from turbulence. For the acoustic field, only spanwise-averaged (2D) data is recorded during the simulation and, hence, only the first approach can be employed.

\begin{figure}[H]
			\centering
			\begin{overpic}[width=0.99\textwidth]{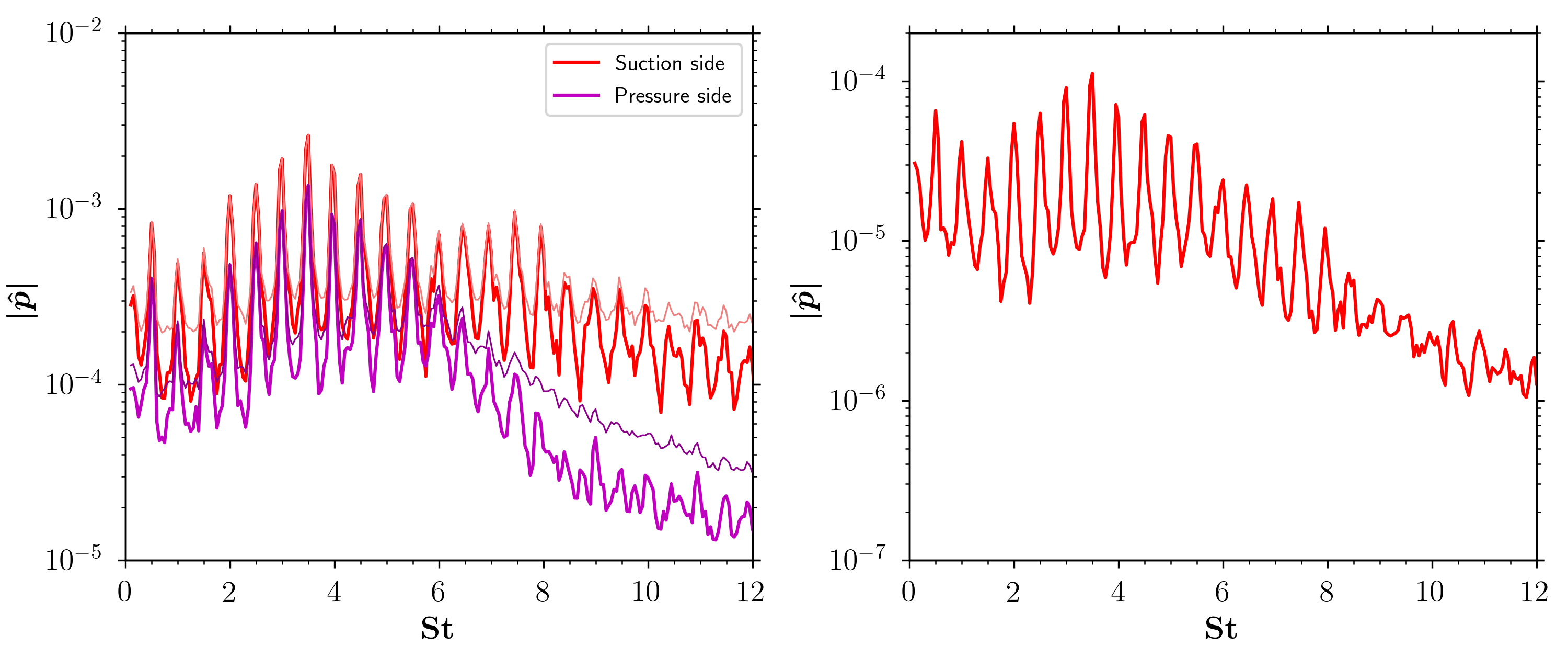}
				\put( 9,37){(a)}
				\put(59,37){(b)}
			\end{overpic}
			\caption{Fourier transform of (a) hydrodynamic pressure signals extracted at the green dots in Fig. \ref{fig:Prms_flow}(a), and (b) nearfield acoustic pressure signal extracted at $x = 1c, y = 1c$, both for $Re = 0.5 \times 10^5$. Thick and thin lines in (a) are obtained with Eqs. (\ref{eq:fft_1}) and (\ref{eq:fft_2}), respectively.}
			\label{fig:fft_50k}
		\end{figure}

		The pressure spectra for the lowest Reynolds number are presented in Figs. \ref{fig:fft_50k}(a) and (b) for hydrodynamic and nearfield acoustic pressure signals, respectively, where it is possible to see the multiple tonal peaks. The main frequency is $St \approx 3.5$ and the secondary tones are equally spaced by $\Delta St \approx 0.5$.  Moreover, all the tonal peaks observed in the spectra are integer multiples of this frequency spacing. Thick and thin lines in Fig. \ref{fig:fft_50k}(a) are obtained with Eqs. (\ref{eq:fft_1}) and (\ref{eq:fft_2}), respectively, and both ways of computing the Fourier transform lead to similar magnitudes of the tonal peaks. Differences are observed mostly on the broadband levels since the spanwise averaging filters out uncorrelated data. Finally, the levels on the pressure side are lower compared to those observed on the suction side in the entire frequency range. This behavior is consistent with the results from Fig. \ref{fig:Prms_flow}. The spectral content on the acoustic field, Fig. \ref{fig:fft_50k}(b), depicts similar trends compared to the hydrodynamic results, but at lower levels due to a decay in the pressure magnitude from the acoustic radiation (note the different $y$-scale levels in the plot).

\begin{figure}[H]
			\centering
			\begin{overpic}[width=0.99\textwidth]{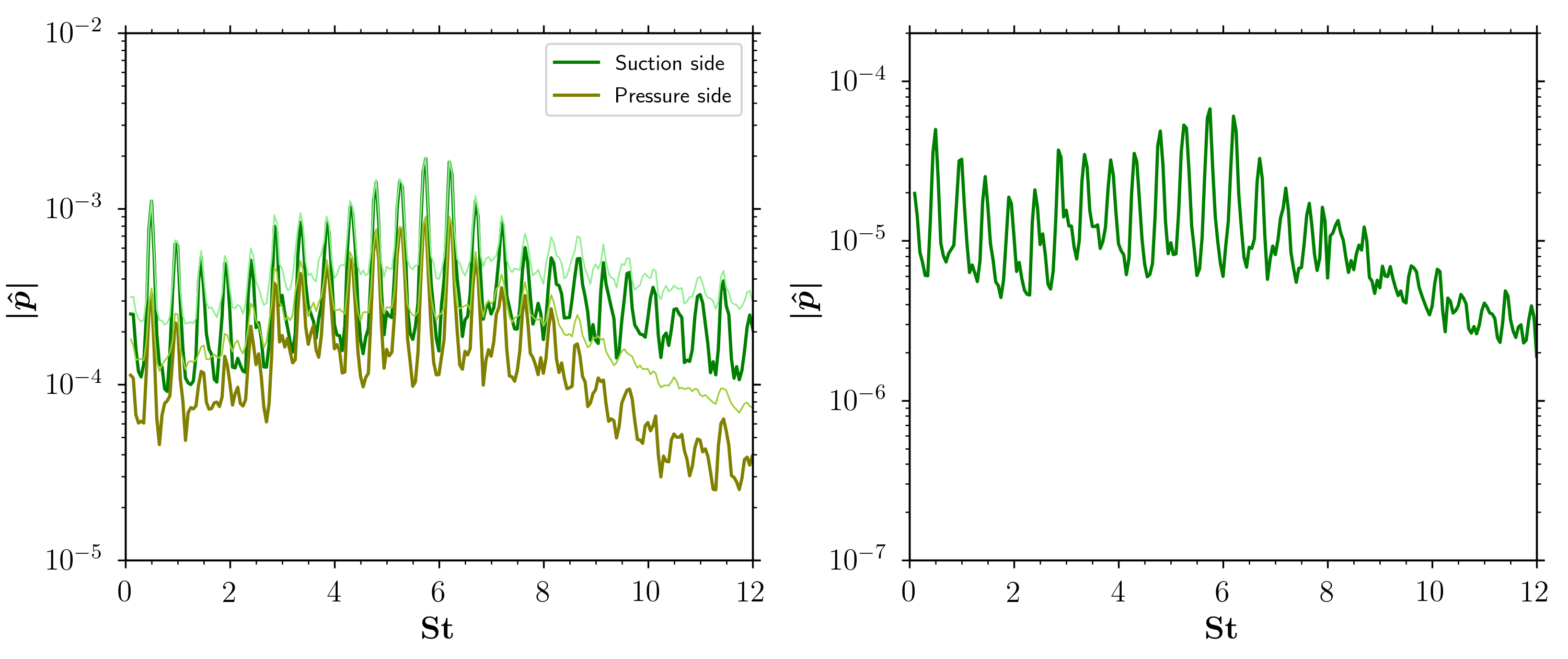}
				\put( 9,37){(a)}
				\put(59,37){(b)}
			\end{overpic}
			\caption{Fourier transform of pressure signals for ${Re = 1 \times 10^5}$ at the (a) hydrodynamic field, extracted at the green dots in Fig. \ref{fig:Prms_flow}(b), and (b) acoustic nearfield, extracted at $x = 1c, y = 1c$.}
			\label{fig:fft_100k}
		\end{figure}

		Results of the $Re = 1 \times 10^5$ simulation are presented in Figs. \ref{fig:fft_100k}(a) and (b) for the hydrodynamic and nearfield acoustic pressure data, respectively. These figures show the main frequency at $St \approx 5.76$ and the multiple tones equally spaced with $\Delta St \approx 0.48$. Again, the bottom side exhibits lower levels of pressure fluctuations compared to the suction side. Also, the two approaches employed in the spectral analysis lead to similar magnitudes for the tonal peaks and the differences are mostly on the broadband component. One important difference compared to the lowest Reynolds number setup is that, for $Re = 1 \times 10^5$, the main peak is observed at a higher frequency since vortex merging seldom occur. This behavior is in agreement with experimental results from \citet[Fig. 5(d)]{Probsting2015_regimes}, who show a similar spectrum for a NACA0012 airfoil at $Re = 1.1 \times 10^5$ and $\alpha = 2.9$ deg. angle of attack. From a visual inspection of the cited figure, six tonal peaks are observed between $2.5 \leq St \leq 5.0$, which gives $\Delta St \approx 0.5$. Moreover, similarly to the present simulations, the previous authors also observe that the main tonal peak shifts to higher frequencies at $Re \approx 1 \times 10^5$.

\begin{figure}[H]
			\centering
			\begin{overpic}[width=0.99\textwidth]{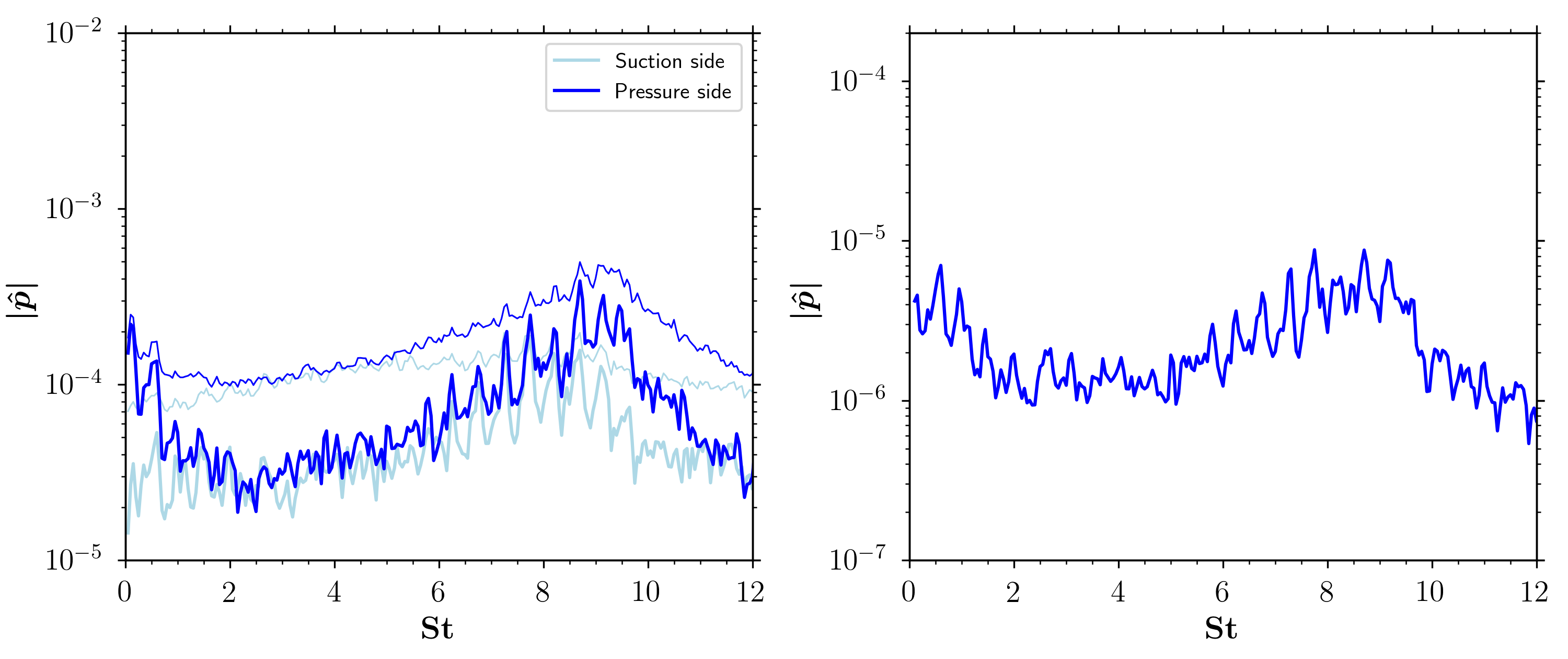}
				\put( 9,37){(a)}
				\put(59,37){(b)}
			\end{overpic}
			\caption{Fourier transform of pressure signals for $Re = 2 \times 10^5$ at the (a) hydrodynamic field, extracted at the green dots in Fig. \ref{fig:Prms_flow}(c), and (b) acoustic nearfield, extracted at $x = 1c, y = 1c$.}
			\label{fig:fft_200k}
		\end{figure}

		Compared to our previous simulated cases, considerable differences can be observed in the spectra when the Reynolds number is increased to $Re = 2 \times 10^5$. Figures \ref{fig:fft_200k}(a) and (b) show the spectra of hydrodynamic and nearfield acoustic pressure, respectively, for this higher Reynolds number setup. In Fig. \ref{fig:fft_200k}(a), it is possible to see that the mid-frequency portion of the spectrum has similar levels for both pressure and suction sides. However, at $St \gtrsim 6$ the pressure side has larger relative levels compared to the suction side. This indicates that, for this higher Reynolds number, the noise sources are switching from the suction to the pressure side. This observation is in agreement with the experimental results by \citet{Probsting2015_regimes}. It is important to mention that, at this Reynolds number, the suction side boundary layer is turbulent at the trailing edge region, differently from the lower Reynolds cases, where laminar-turbulent transition occurred in an intermittent fashion at the trailing edge. Nonetheless, it is possible to see that all curves show tonal peaks ranging from $5.78<St<9.18$ with a $\Delta St \approx 0.48$. The acoustic spectrum in Fig. \ref{fig:fft_200k}(b) also presents tonal peaks at the same frequencies observed in the hydrodynamic field. One important comment is that the magnitude in the pressure spectra for this particular Reynolds number are one order of magnitude lower compared to the previous cases. Furthermore, the peak heights are also lower, which indicates that the broadband content due to turbulent transition becomes more relevant.

\begin{figure}[H]
			\centering
			\begin{overpic}[width=0.99\textwidth]{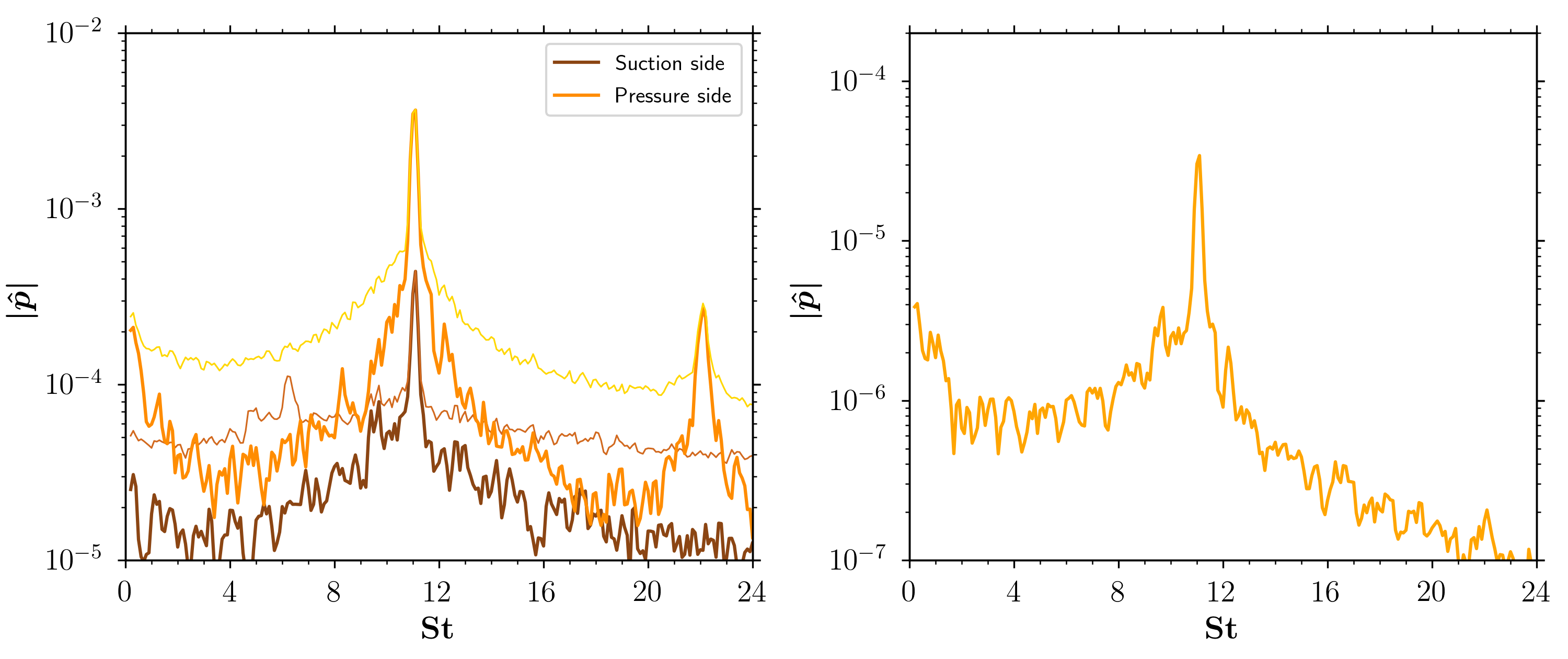}
				\put( 9,37){(a)}
				\put(59,37){(b)}
			\end{overpic}
			\caption{Fourier transform of pressure signals for ${Re = 4 \times 10^5}$ at the (a) hydrodynamic field, extracted at the green dots in Fig. \ref{fig:Prms_flow}(d), and (b) acoustic nearfield, extracted at $x = 1c, y = 1c$.}
			\label{fig:fft_400k}
		\end{figure}

		The highest Reynolds number investigated presents different trends compared to the previous cases in terms of spectral content. The Fourier transform of the hydrodynamic pressure data is presented in Fig. \ref{fig:fft_400k}(a), where it is possible to see a more prominent tonal peak at $St = 11$ and its harmonic at $St = 22$. For this case, the pressure side levels are one order of magnitude higher compared to those computed on the suction side, which may indicate that the latter does not play an important role in the acoustic emission. Figure \ref{fig:fft_400k}(b) presents the pressure spectrum computed in the acoustic nearfield and it depicts the tonal peak at $St = 11.0$, and its almost imperceptible first harmonic at $St = 22.0$.


    \clearpage
\subsection{Noise generation mechanisms}

	Trailing-edge noise generation by airfoils depends on the spanwise coherence of flow structures and their subsequent acoustic scattering. The scattered pressure field follows a $1/R^3$ dependence, where $R$ is the distance from the sources to the trailing edge. Hence, flow disturbances are most relevant in the vicinity of the trailing edge \cite{FWHall1970} and, as discussed by \citet{sanoprf2019}, spanwise-correlated flow structures are more efficient in terms of trailing-edge noise radiation. In this regard, the pressure field is initially filtered by a spanwise averaging in order to remove uncorrelated content that does not radiate efficiently while still retaining the sources aligned with the trailing edge. Then, the 2D flow field is further decomposed by the POD to identify important spanwise acoustic sources and their distribution along the airfoil.

	Although the POD is computed using only the pressure fluctuations, other norms were also investigated for the construction of the covariance matrix, with results leading to the same conclusions. The present methodology allows the identification of dominant flow features, shown as the spatial modes, and their respective frequency spectra, presented as Fourier transforms of the temporal modes. These are shown qualitatively with the purpose of visualization and the levels are omitted. Following the SVD decomposition in Sec. \ref{sec:pod_form}, they must satisfy $\boldsymbol{\phi}_i^T \mathbf{W}\boldsymbol{\phi}_j = \boldsymbol{\delta_{ij}}$ and $a_i a_i = \sigma_{i}^2$, where $\delta_{ij}$ is the Kronecker delta function and $\sigma_i$ represents the singular values of the $i^{th}$ mode.

\begin{figure}
		\centering
		\begin{overpic}[width=0.95\textwidth]{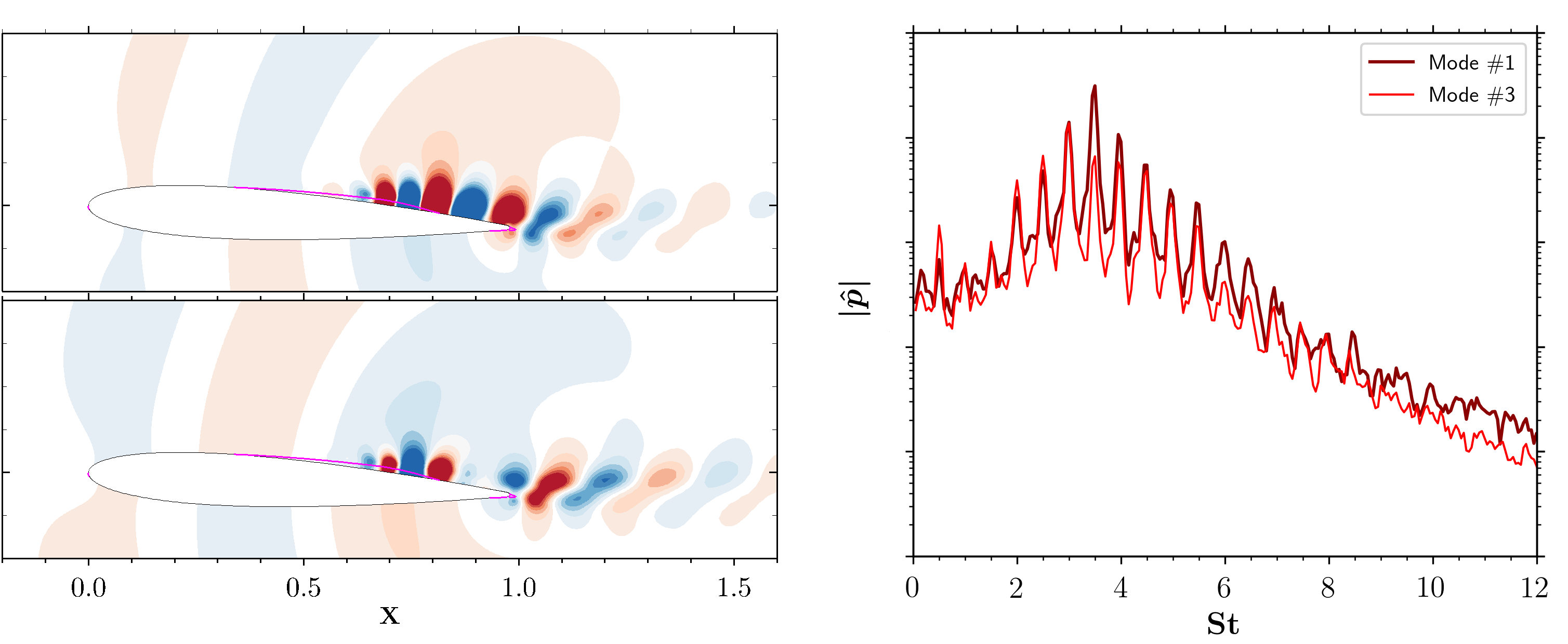}
			\put(1,36){(a)}
			\put(1,19){(b)}
			\put(59,36){(c)}
		\end{overpic}
		\caption{POD analysis for $Re = 0.5\times 10^5$ in terms of pressure spatial modes (a) \#1 and (b) \#3, and (c) Fourier-transformed temporal modes.}
		\label{fig:pod_2D_modes_Re50k}
	\end{figure}

	The decomposition results for $Re = 0.5 \times 10^5$ in terms of spatial modes \#1 and \#3 are presented in Figs. \ref{fig:pod_2D_modes_Re50k}(a,b). The even modes \#2 and \#4 are paired with their odd counterparts and, hence, depict similar spatial and frequency characteristics. It is possible to see that the dominant flow structures appear downstream the separation bubble, indicated by the magenta line, along the suction side. While mode \#1 has a higher magnitude along the airfoil surface, mode \#3 depicts strong fluctuations both on the reattachment region and along the wake. Acoustic waves can be seen radiating from the trailing edge in the same figures. The Fourier-transformed temporal modes are presented in Fig. \ref{fig:pod_2D_modes_Re50k}(c), where the multiple peaks are observed, in agreement with Fig. \ref{fig:fft_50k}. The first two pairs of POD modes account for more than 35\% of the data variance. Thus, these modes illustrate the main characteristics of the flow.

\begin{figure}
	\centering
	\begin{overpic}[width=0.95\textwidth]{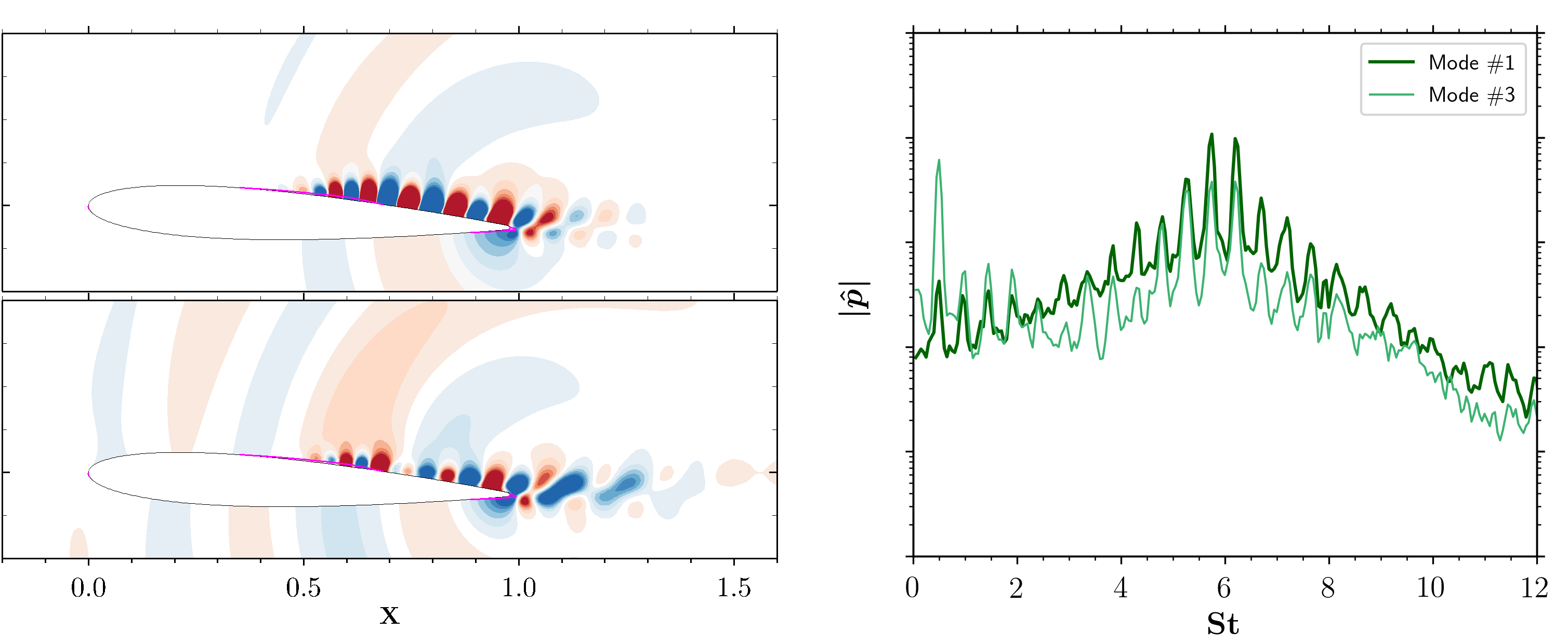}
		\put(1,36){(a)}
		\put(1,19){(b)}
		\put(59,36){(c)}
	\end{overpic}
	\caption{POD analysis for $Re = 1\times 10^5$ in terms of pressure spatial modes (a) \#1 and (b) \#3, and (c) Fourier-transformed temporal modes.}
	\label{fig:pod_2D_modes_Re100k}
	\end{figure}

	Similar features are observed for $Re = 1 \times 10^5$ as shown in Figs. \ref{fig:pod_2D_modes_Re100k}(a,b) in terms of spatial modes \#1 and \#3, computed for pressure fluctuations. In Fig. \ref{fig:pod_2D_modes_Re100k}(c), the Fourier-transformed temporal modes are shown. Similarly to the previous case, the dominant flow features are observed on the suction side and along the wake. The temporal modes present all the multiple peaks observed in Fig. \ref{fig:fft_100k} and the first two mode pairs account for 24\% of the data variance.

\begin{figure}
	\centering
	\begin{overpic}[width=0.95\textwidth]{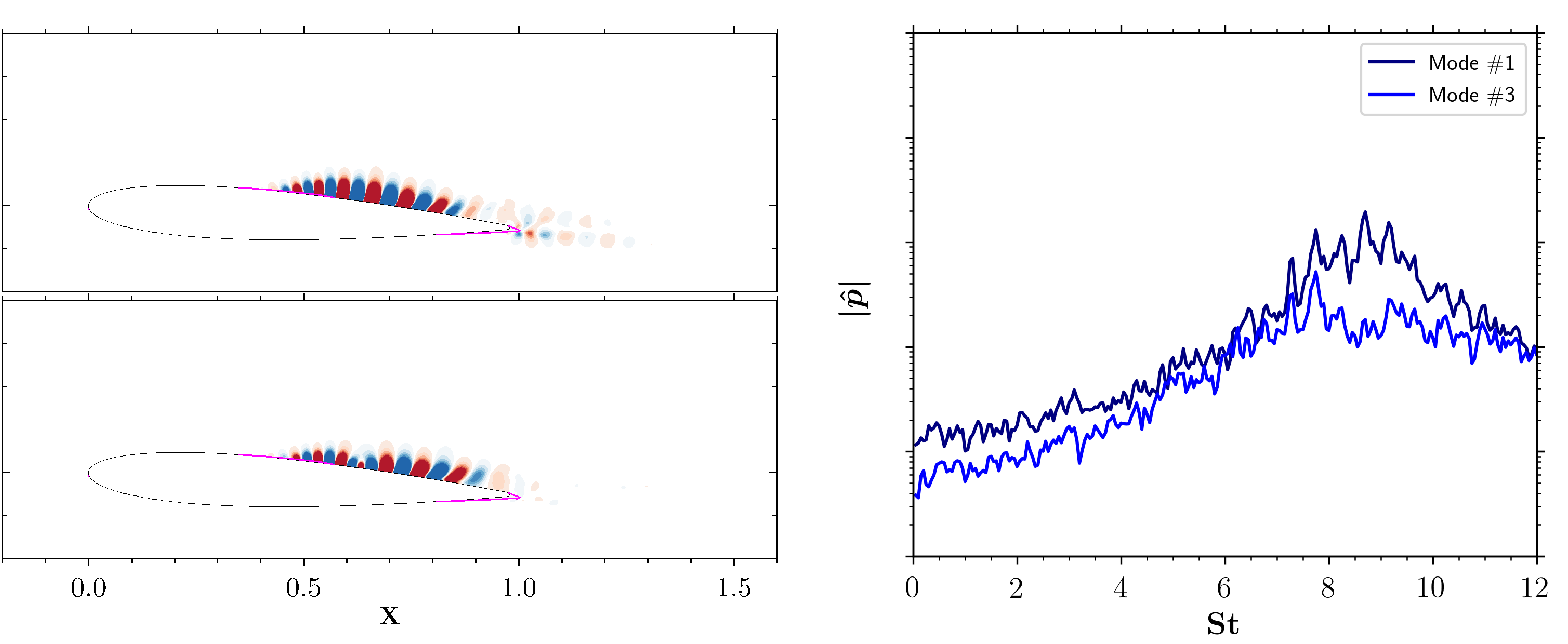}
		\put(1,36){(a)}
		\put(1,19){(b)}
		\put(59,36){(c)}
	\end{overpic}
	\caption{POD analysis for $Re = 2\times 10^5$ in terms of pressure spatial modes (a) \#1 and (b) \#3, and (c) Fourier-transformed temporal modes.}
	\label{fig:pod_2D_modes_Re200k}
	\end{figure}

	For $Re = 2 \times 10^5$, the spatial modes presented in Figs. \ref{fig:pod_2D_modes_Re200k}(a,b) show that the spanwise coherent structures also appear downstream the laminar separation bubble, but their spatial support has a faster decay towards the trailing edge. Furthermore, the first mode depicts flow instabilities along the wake, mostly on the pressure side. This corroborates with the larger levels of fluctuations observed below the airfoil in Fig. \ref{fig:fft_200k}. The temporal modes are presented in Fig. \ref{fig:pod_2D_modes_Re200k}(c) and the multiple tones are still observed, despite the more pronounced broadband levels when compared to the previous flow configurations analyzed. The first two mode pairs add up to nearly 50\% of the total variance.

\begin{figure}
	\centering
	\begin{overpic}[width=0.95\textwidth]{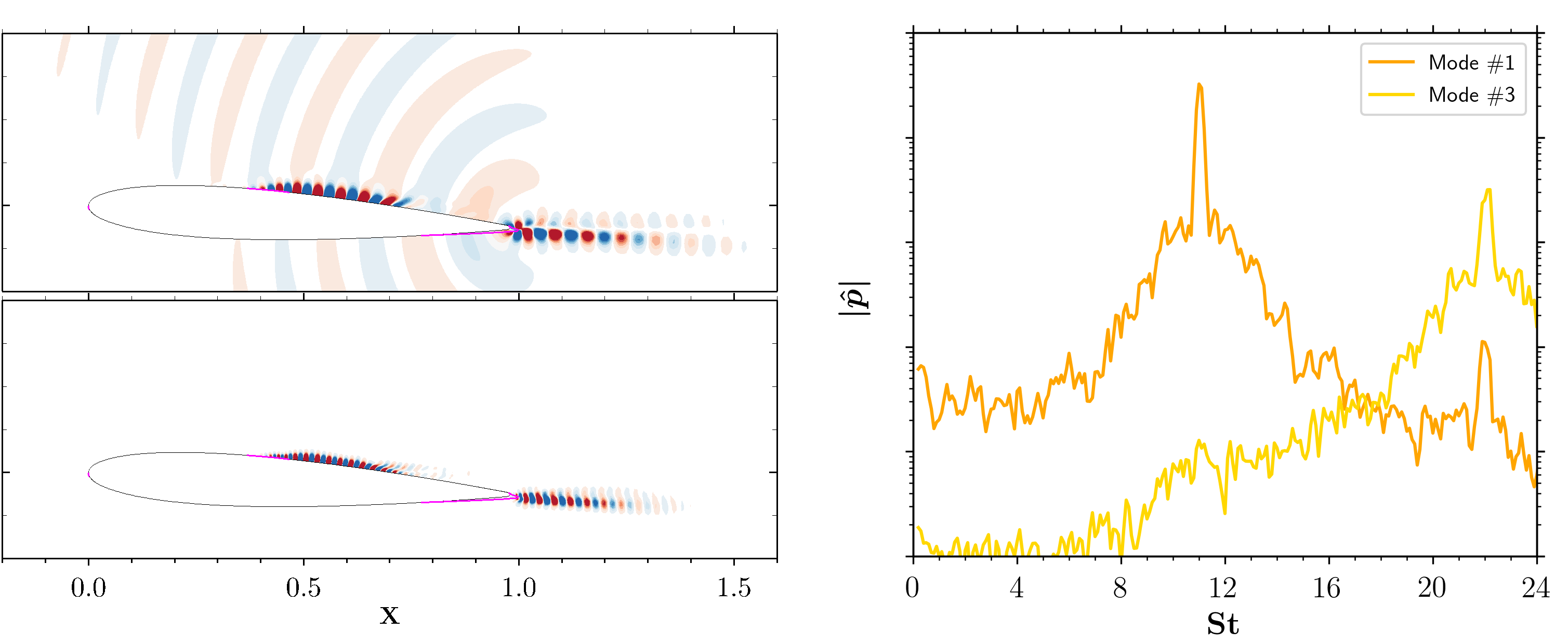}
		\put(1,36){(a)}
		\put(1,19){(b)}
		\put(59,36){(c)}
	\end{overpic}
	\caption{POD analysis for $Re = 4\times 10^5$ in terms of pressure spatial modes (a) \#1 and (b) \#3, and (c) Fourier-transformed temporal modes.}
	\label{fig:pod_2D_modes_Re400k}
	\end{figure}

	For the highest Reynolds number, $Re = 4 \times 10^5$, the spatial modes depicted in Figs. \ref{fig:pod_2D_modes_Re400k}(a,b) show that the flow patterns from the suction side lose their coherence downstream of the tiny separation bubble. These spatial modes also highlight the importance of the pressure side and aerodynamic wake in the trailing-edge noise generation. The temporal modes presented in Fig. \ref{fig:pod_2D_modes_Re400k} depict the dominant tonal frequency at $St = 11.0$ and its harmonic at $St = 22.0$. For this case, the variance of the first two mode pairs results in 80\% of the flow energy.

	One important comment can be made regarding the appearance of laminar instabilities on the suction side. Such flow patterns appear with the same frequencies of the wake instabilities from the pressure side, on the trailing edge. This may be triggered by the acoustic feedback loop mechanism discussed in literature \cite{Tam1974,Arbey1983,Lowson1994,Desquesnes2007,Plogmann2013} or by changes in circulation due to vortex shedding which may alter the instantaneous angle of attack.

    \clearpage
	\subsection{Flow intermittency}

	The intermittency of coherent structures plays an important role in the amplitude modulation and noise generation at the trailing edge \cite{Desquesnes2007, Probsting2014, Padois2016, Sanjose2018, Ricciardi2019_tones, Yakhina2020, ricciardi2022JFM}. Thus, time-dependent analyses using the continuous Morlet wavelet transform (CWT) and the two-point, one-time autocovariance of pressure fluctuations along the span, defined by Eq. (\ref{eq:correlation}), identify time instants when spanwise coherent structures reach the trailing edge. Similar covariance plots were also computed in terms of velocity fluctuations (not shown for brevity) and led to the same observations. To illustrate this behavior, instantaneous flow solutions are presented in terms of $\omega_z$ vorticity as red-blue contours. The measure of entropy $\sfrac{p}{\rho^{\gamma}} - \sfrac{p_\infty}{\rho_\infty^{\gamma}} = 0.1\%$ is used to highlight spots of vorticity and it is depicted as a black line, while magenta lines indicate the LSB based on the time-averaged reversed flow, $\bar{u} < 0$. The time signals are extracted at the same locations discussed in Sec. \ref{sec:SpectralAnalysis}. For the lower (higher) Reynolds numbers, signals are computed at the green dots depicted in Fig. \ref{fig:Prms_flow}, positioned on the suction (pressure) side. The flow snapshots are extracted from movie \#2, in the Supplemental Material, which shows the spanwise averaged flow dynamics.

	Results of the intermittency analysis are presented in Figs. \ref{fig:wavelet_50k} and \ref{fig:wavelet_100k} for $Re = 0.5\times 10^5$ and $1 \times 10^5$, respectively. Examples of coherent structures near the trailing edge are presented in sub-figures (a), while uncorrelated turbulent eddies are shown in sub-figures (b). In the autocovariance plot, presented in sub-figures (c), the coherent flow structures correspond to dark lines that extend the entire spanwise direction, while the uncorrelated turbulence coincides with faint lines or the white color.

	Analyzing the autocovariance in conjunction with the CWT scalogram, shown in sub-figures (d), it is possible to see that specific discrete frequencies are excited when the coherent structures pass by the trailing edge. However, each one of the tones is not necessarily excited at all times, indicating that the multiple tonal peaks are intermittent and that the main frequency shown by the red spots may change to one of the secondary peaks. Despite this, for the lowest Reynolds number, the tone at $St \approx 3.5$ is the most prominent for almost the entire time history analyzed. Increasing the Reynolds number to $Re = 1\times10^5$, it is not possible to see a single more prominent peak during the  simulation period. In this case, the instantaneous main frequency alternates between $4.78 \le St \le 6.20$ with increments of $\Delta St \approx 0.48$, i.e., the first tonal peak frequency observed in the Fourier transform in Fig \ref{fig:fft_100k}. A similar frequency switch is observed in the 2D simulations from \citet{Ricciardi2019_tones} and the experimental measurements from \citet{Yakhina2020}.
	\begin{figure}
		\centering
		\begin{overpic}[width=0.99\textwidth]{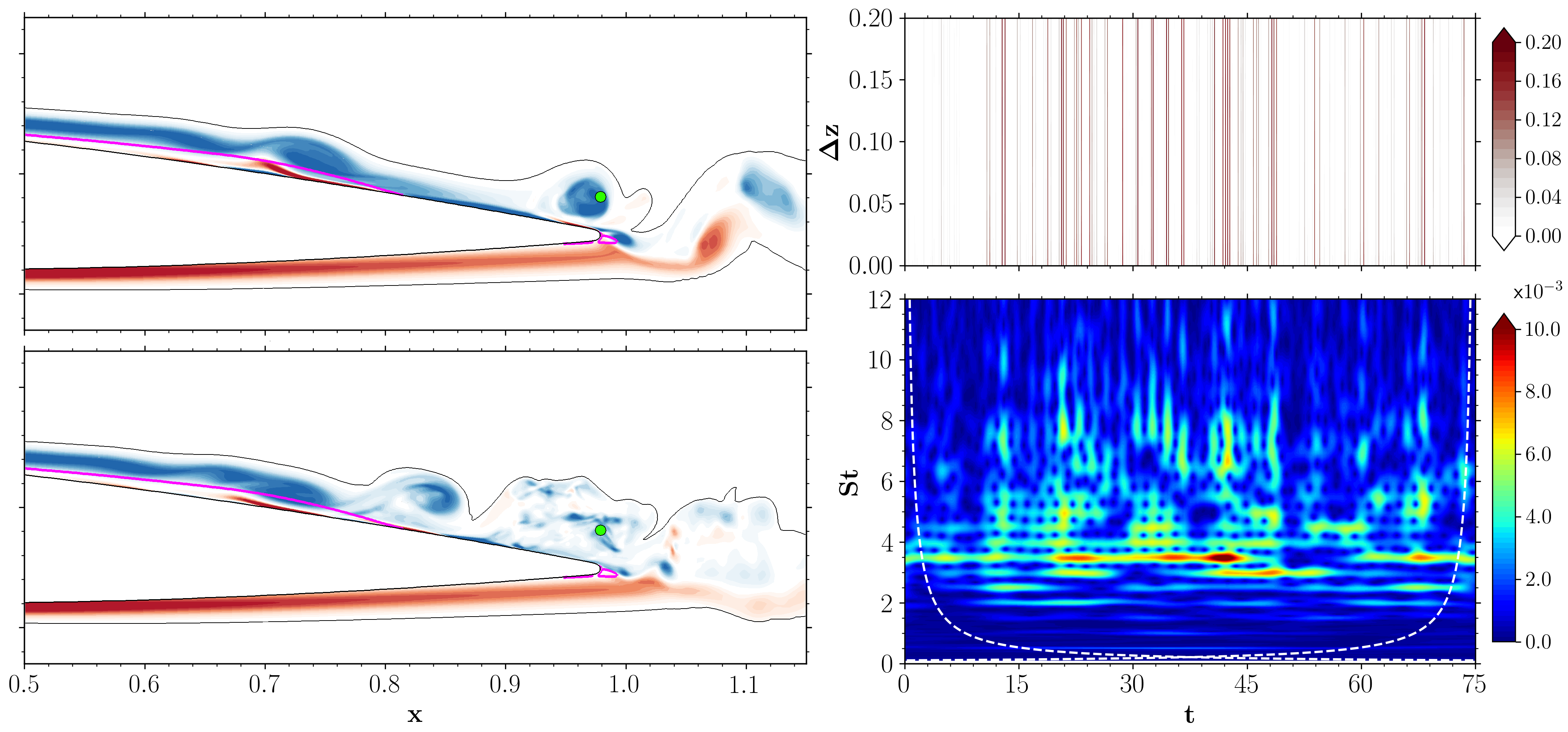}
			\put(2.50,42.75){(a) Coherent structure, $t = 20.82$}
			\put(2.50,21.50){(b) Uncorrelated eddies, $t = 21.78$}
			\put(58.5,43.00){(c)}
			\put(58.5,24.75){\white{(d)}}
		\end{overpic}
		\caption{Time-dependent analysis of pressure signals acquired at the green dot for ${Re = 0.5 \times 10^5}$ presented as (a,b) flow snapshots, (c) two-point, one-time autocovariance, and (d) CWT.}
		\label{fig:wavelet_50k}
	\end{figure}
	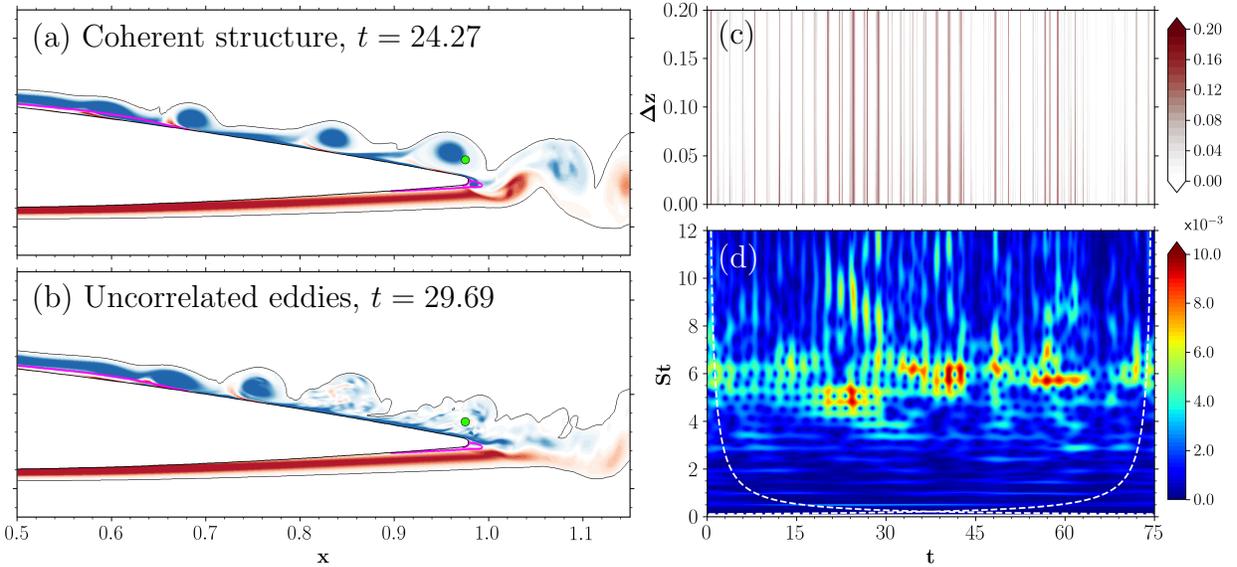
\begin{figure}
		\centering
		\begin{overpic}[width=0.99\textwidth]{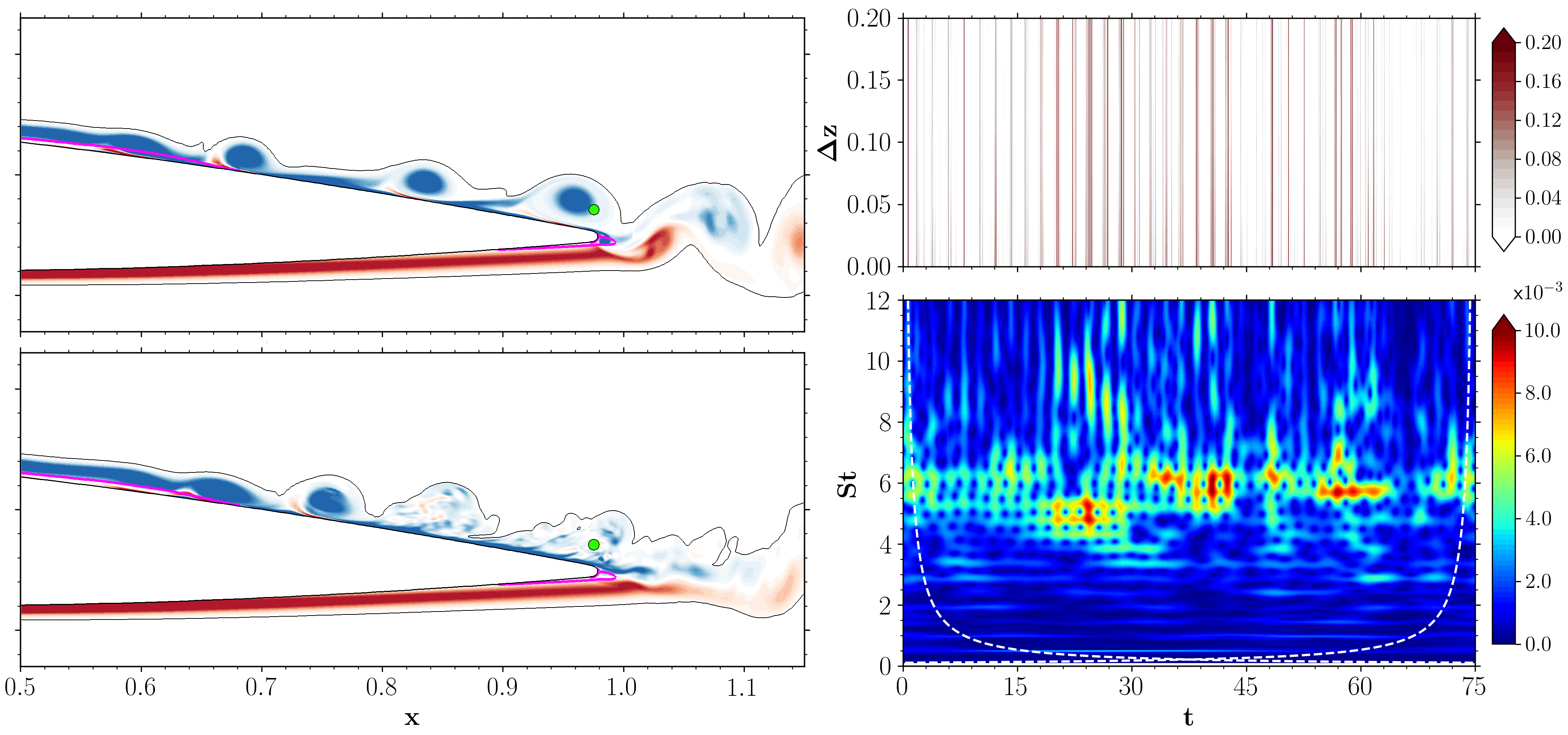}
			\put(2.50,42.75){(a) Coherent structure, $t = 24.27$}
			\put(2.50,21.50){(b) Uncorrelated eddies, $t = 29.69$}
			\put(58.5,43.00){(c)}
			\put(58.5,24.75){\white{(d)}}
		\end{overpic}
		\caption{Time-dependent analysis of pressure signals acquired at the green dot for ${Re = 1 \times 10^5}$ presented as (a,b) flow snapshots, (c) two-point, one-time autocovariance, and (d) CWT.}
		\label{fig:wavelet_100k}
	\end{figure}

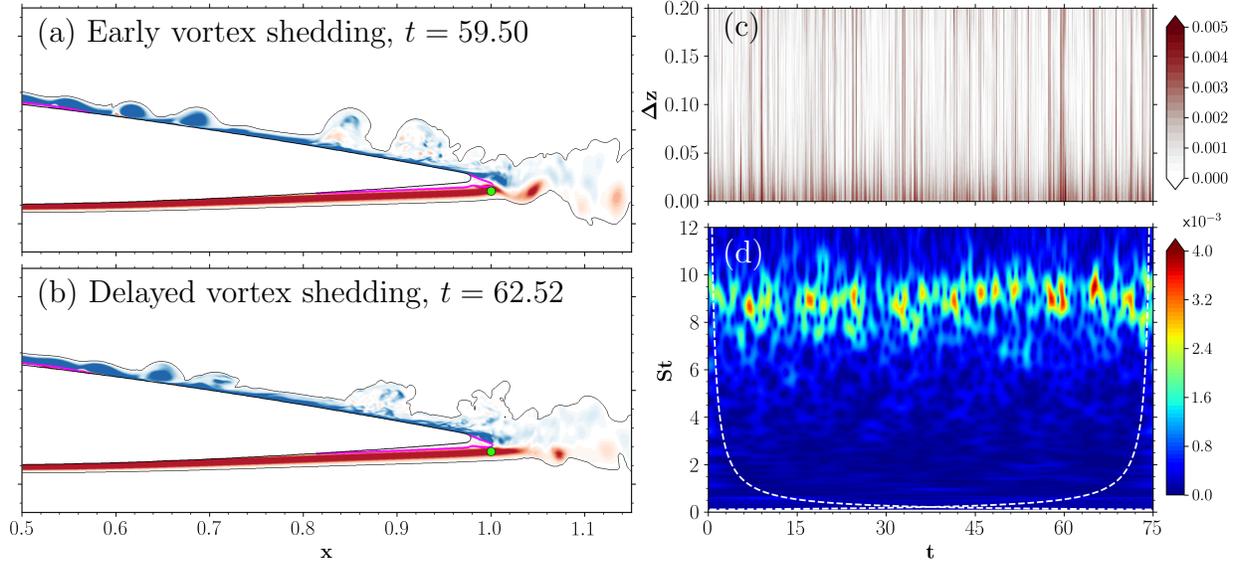
\begin{figure}
		\centering
		\begin{overpic}[width=0.99\textwidth]{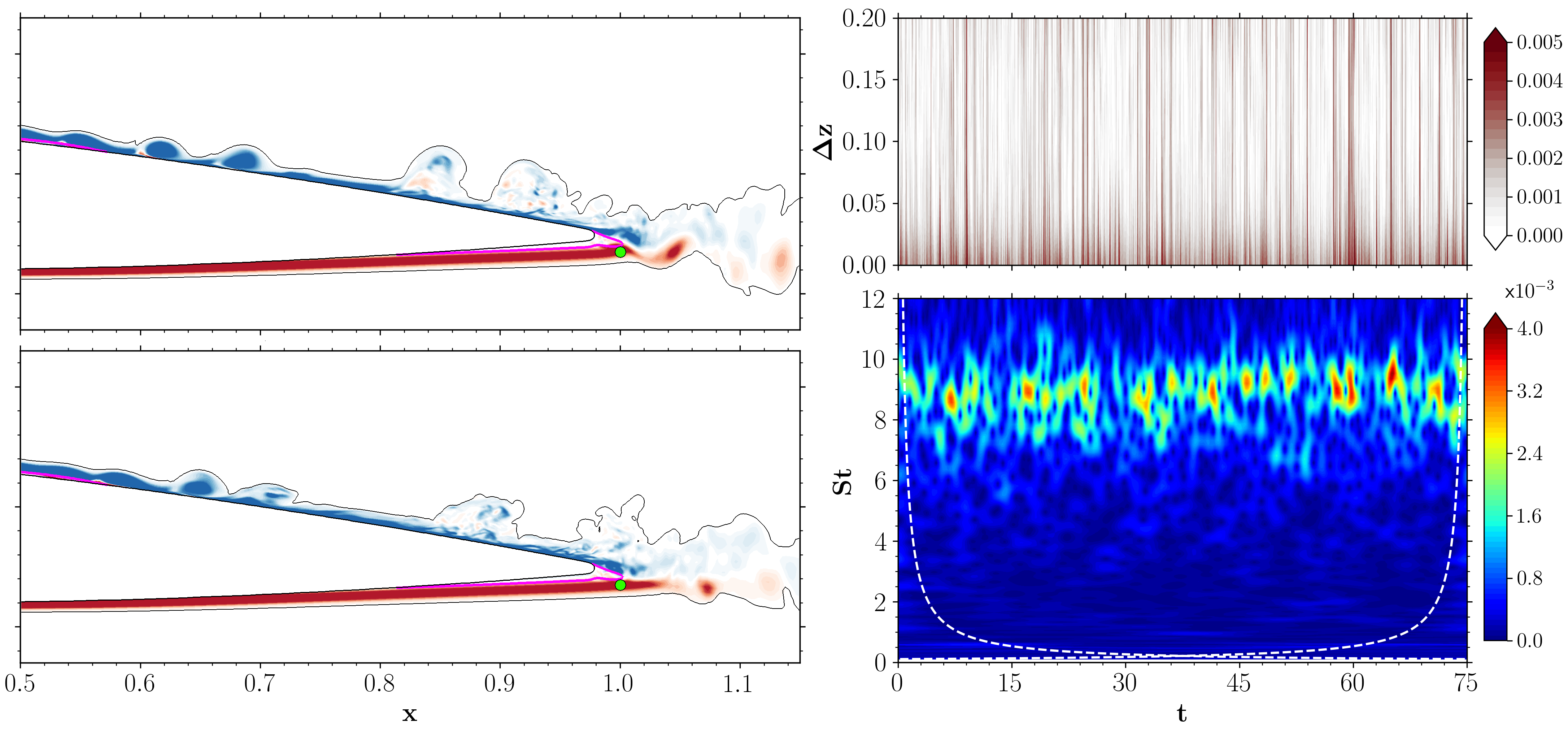}
			\put(2.50,42.75){(a) Early vortex shedding, $t = 59.50$}
			\put(2.50,21.50){(b) Delayed vortex shedding, $t = 62.52$}
			\put(58.5,43.25){(c)}
			\put(58.5,24.75){\white{(d)}}
		\end{overpic}
		\caption{Time-dependent analysis of pressure signals acquired at the green dot for ${Re = 2 \times 10^5}$ presented as (a,b) flow snapshots, (c) two-point, one-time autocovariance, and (d) CWT.}
		\label{fig:wavelet_200k}
	\end{figure}

	When the Reynolds number is increased to $Re = 2 \times 10^5$, the flow snapshots presented in Figs. \ref{fig:wavelet_200k}(a,b) show that the suction side only exhibits turbulent packets reaching the trailing edge. For this case, coherent structures are no longer observed on this side of the airfoil. However, spanwise coherence is observed on the pressure side, downstream the trailing edge. In Fig. \ref{fig:wavelet_200k}(a), one can observe a time instant when a vortex is shed at the trailing edge while Fig. \ref{fig:wavelet_200k}(b) shows an instant when the shear-layer extends further downstream, without the roll-up. Such time instants are identified based on the autocovariance and CWT scalograms presented in Figs. \ref{fig:wavelet_200k}(c) and (d), respectively. Another important comment is that, in opposition to the lower Reynolds numbers, both the CWT scalograms and the autocovariances show that the flow does not have an organized motion. The tonal peaks are related to an erratic vortex shedding rather than a quasi-periodic motion. Still, such events happen at particular discrete frequencies based on the Fourier transform presented in Fig. \ref{fig:fft_200k}. Finally, it is observed that interactions between the suction and pressure sides may shift the shedding location and push the vortices further downstream. Hence, the truncation and rounding of the trailing edge may be important for this moderate Reynolds number flow since noise sources can switch between suction and pressure sides.

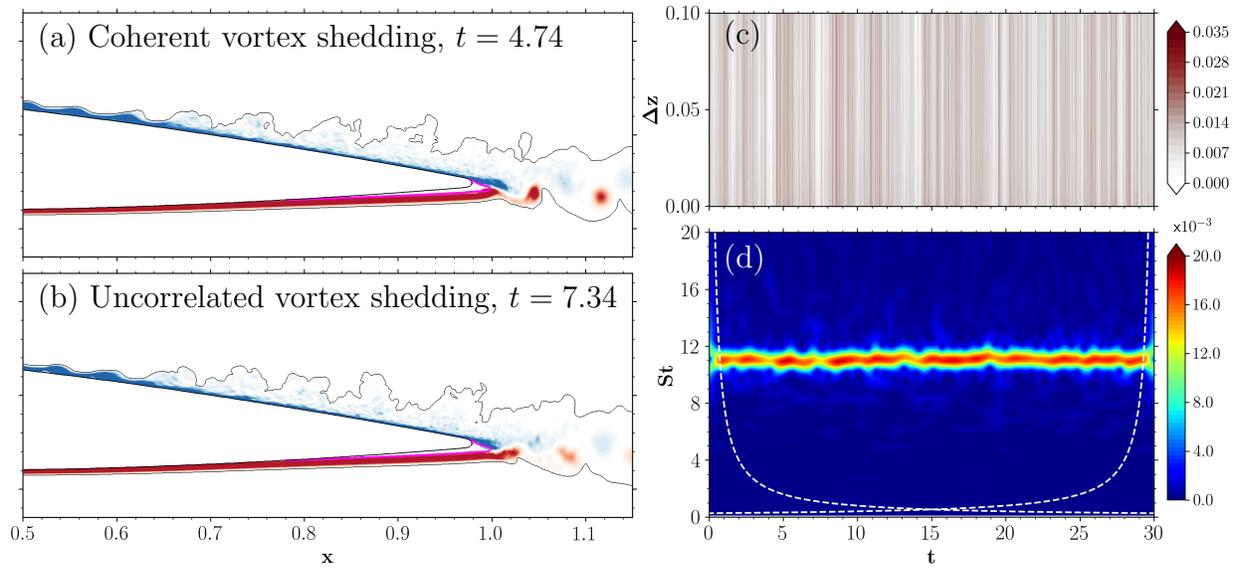
\begin{figure}
		\centering
		\begin{overpic}[width=0.99\textwidth]{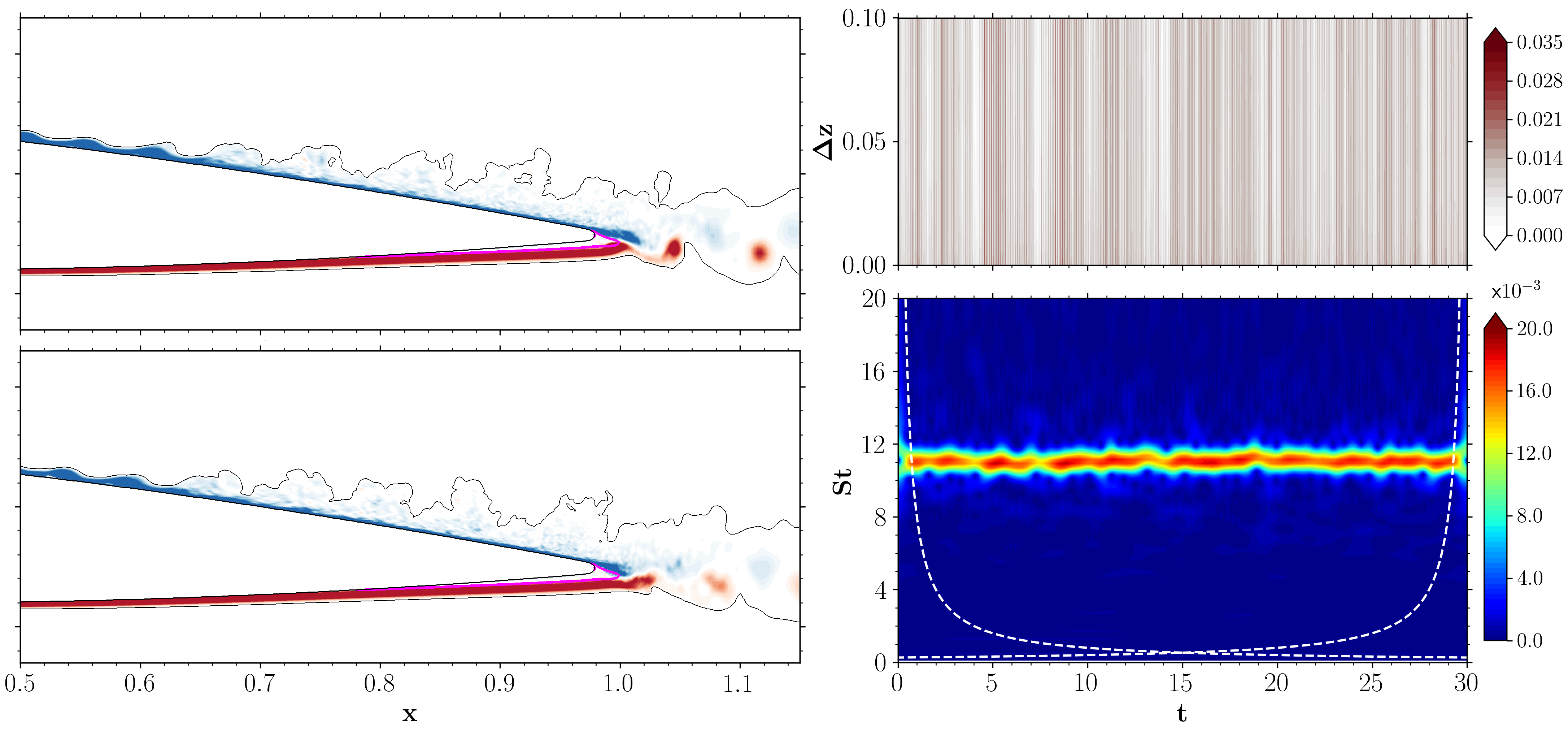}
			\put(2.50,42.75){(a) Coherent vortex shedding, $t = 4.74$}
			\put(2.50,21.50){(b) Uncorrelated vortex shedding, $t = 7.34$}
			\put(58.5,43.00){(c)}
			\put(58.5,24.75){\white{(d)}}
		\end{overpic}
		\caption{Time-dependent analysis of pressure signals acquired at the green dot for ${Re = 4 \times 10^5}$ presented as (a,b) flow snapshots, (c) two-point, one-time autocovariance, and (d) CWT.}
		\label{fig:wavelet_400k}
	\end{figure}

	Figures \ref{fig:wavelet_400k}(a,b) show snapshots of the highest Reynolds number flow analyzed and it is possible to see the vortex shedding from the pressure side. The autocovariance for this flow is presented in Fig. \ref{fig:wavelet_400k}(c), where the well defined dynamics is characterized by the periodic shedding of more organized flow structures. Differently from the previous cases, only slight variations in the autocovariance magnitude can be observed during the simulation temporal window. This is better described by the CWT, where the large wavelet coefficients in the scalogram presented in Fig. \ref{fig:wavelet_400k}(d) indicate an almost monochromatic wave. For this case, intermittent events are no longer observed.

    \clearpage
	\subsection{Pressure side bubble}

\begin{figure}
		\centering
		\begin{overpic}[width=0.99\textwidth]{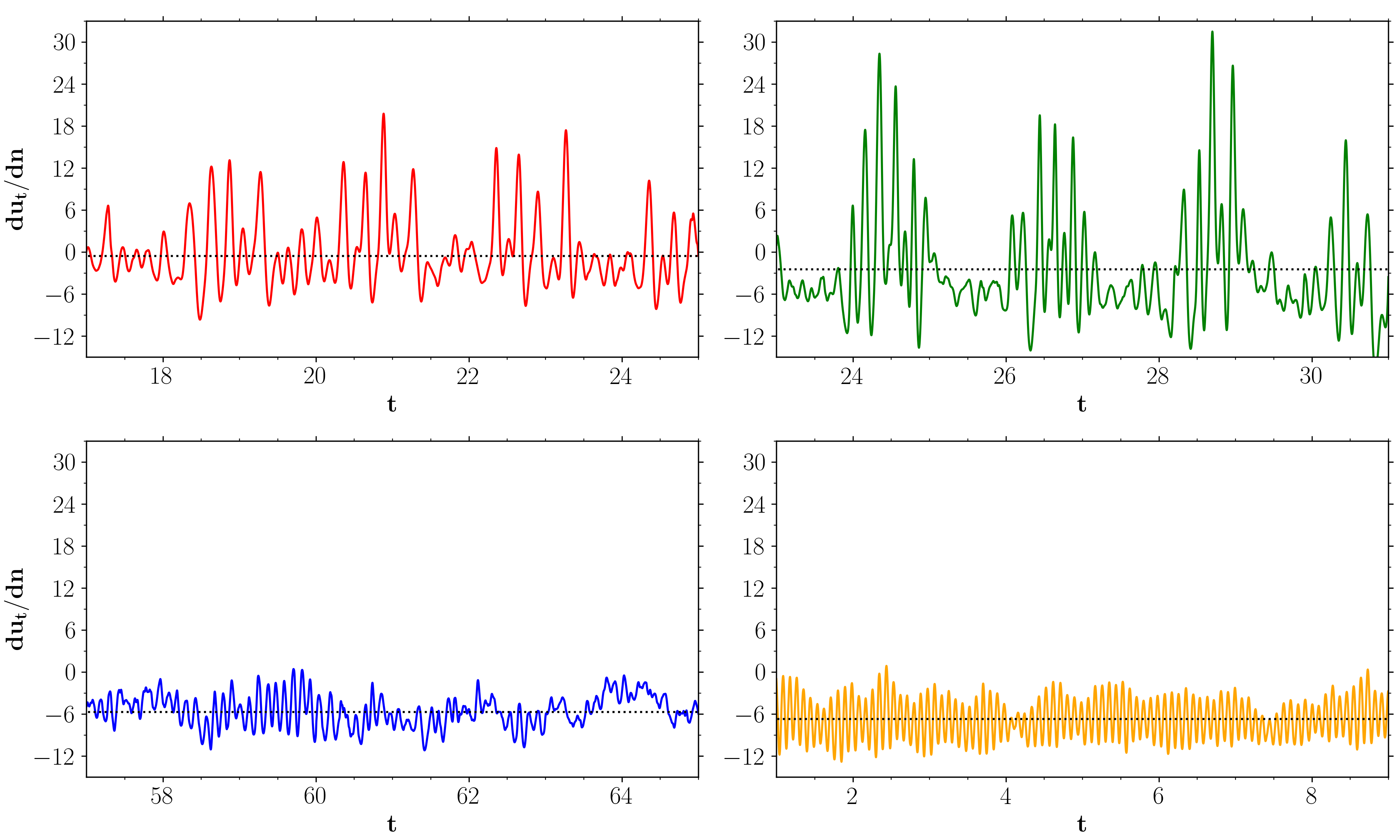}
			\put(0,30){(a) $Re = 0.5 \times 10^5$}
			\put(52,30){(b) $Re = 1 \times 10^5$}
			\put(0,0){(c) $Re = 2 \times 10^5$}
			\put(52,0){(d) $Re = 4 \times 10^5$}
		\end{overpic}
		\caption{Derivative of tangential velocity along the wall-normal direction $\mbox{du}_{\mbox{\scriptsize{t}}}\mbox{/dn}$ at $x = 0.95$ on the pressure side. The dotted line represents the averaged value.}
		\label{fig:probe_PSbubble}
	\end{figure}
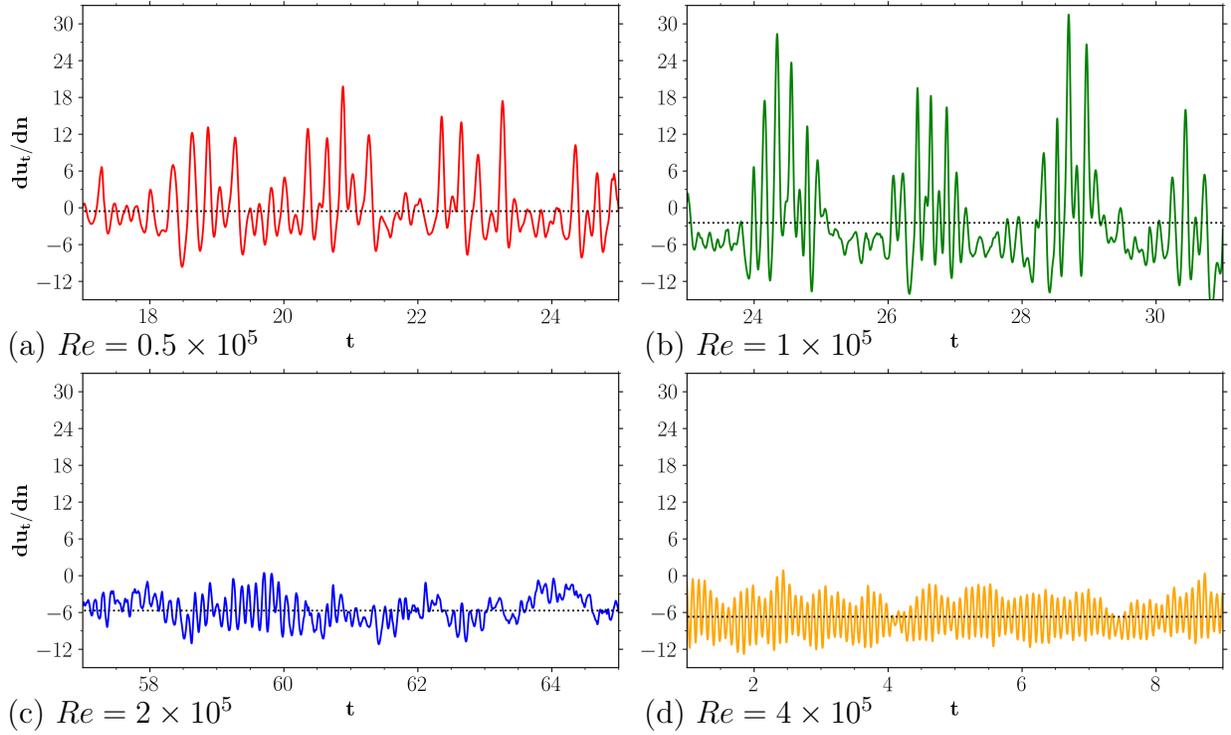

	An important aspect observed in the present simulations concerns the presence of the pressure side bubble. For lower Reynolds numbers, where the suction side dominates the flow dynamics, it is observed that the pressure side bubbles are intermittent and dependent on the advection of coherent structures at the trailing edge. In order to understand this intermittency, the temporal signal of the tangential velocity wall-normal derivative $\mbox{du}_{\mbox{\scriptsize{t}}}\mbox{/dn}$ on the pressure side, at $x=0.95$, is presented in Figs. \ref{fig:probe_PSbubble}(a) and (b) for $Re = 0.5$ and $1 \times 10^5$, respectively. The derivative is computed using the wall-normal direction positive when pointing away from the surface and it indicates whether the flow is attached $(\mbox{du}_{\mbox{\scriptsize{t}}}\mbox{/dn} > 0)$ or recirculating $(\mbox{du}_{\mbox{\scriptsize{t}}}\mbox{/dn} < 0)$ in an instantaneous sense as opposed to the time-averaged value shown by the dotted line. The positive spikes in the signal are observed when the coherent structures from the suction side leave the trailing edge.

	When the Reynolds number increases, the suction side coherent structures are no longer observed due to turbulence. This has a direct influence on the pressure side bubble as presented in Figs. \ref{fig:probe_PSbubble}(c) and (d) for $Re = 2$ and $4 \times 10^5$, respectively. In these figures, it is possible to see that the spikes vanish and the velocity derivative fluctuations near the wall are reduced. Furthermore, the signal has negative values of $\mbox{du}_{\mbox{\scriptsize{t}}}\mbox{/dn}$ for the time majority, indicating that the LSB is permanent rather than intermittent. Finally, it is possible to see that the $Re = 2 \times 10^5$ flow presents a rather erratic motion of the bubble which impacts the vortex shedding, as shown in Fig. \ref{fig:wavelet_200k}. On the other hand, the $Re = 4 \times 10^5$ case has a more organized bubble behavior which, in turn, leads to the efficient noise generation observed in tonal peak of Fig. \ref{fig:wavelet_400k}.

    \clearpage
\section{Conclusions}

The Reynolds number influence on the noise emission from a NACA0012 airfoil is investigated by means of wall-resolved large eddy simulations. The airfoil is fixed at an angle of attack of $\alpha = 3$ deg. and the freestream Mach number is set as $M = 0.3$. Four different simulations are performed at moderate Reynolds numbers $Re = 0.5$, 1, 2 and $4 \times 10^5$ to investigate the mechanisms of noise generation and the switch from suction to pressure side dominated dynamics. It has been shown that laminar separation bubbles are responsible for shedding of spanwise-coherent structures which, in turn, trigger acoustic waves at the trailing edge. Current results qualitatively agree with experiments and other simulations reported in the literature. Here, we provide additional insights onto the different regimes of airfoil tonal noise generation. For instance, the analyses of spanwise flow coherence and intermittency identify different flow patterns for each one of the configurations analyzed. Thus, it is possible to understand the associated noise generation mechanisms and how they affect the noise spectrum.

The major differences in the analyzed flows are observed in terms of mean values of streamwise velocity, where it is possible to see that the lower Reynolds number flows present more pronounced recirculation bubbles on the suction side. On the other hand, increasing the Reynolds number not only reduces the suction side recirculation but also results in a more prominent bubble on the pressure side. The characteristic length of the detached flow, in turn, affects the size of the coherent structures and the position where they originate.

The vortex dynamics downstream of the reattachment point is of utmost importance for the noise generation process, which depends on the spanwise coherence of the sources of sound aligned with the trailing edge. In this regard, coherent structures are identified by visualization of spanwise vortical structures and proper orthogonal decomposition. The locations where they are spanwise coherent is depicted by POD spatial modes and by RMS quantities. For the lower Reynolds numbers, $Re = 0.5$ and $1 \times 10^5$, organized flow patterns are observed on the suction side, including the vicinity of the trailing edge. When increasing the Reynolds numbers to $Re = 2$ and $4 \times 10^5$, coherent vortex shedding is only observed at the trailing edge, arising from the pressure side.

Similar flow dynamics are observed between the configurations where the suction side is responsible for the main flow features. For instance, the flows with Reynolds numbers $Re = 0.5$ and $1 \times 10^5$ are characterized by either coherent structures or turbulent packets being advected along the suction side. Also, the pressure side bubble is only defined in an averaged sense since the flow detachment on the airfoil bottom side is also intermittent and dependent on the flow dynamics at the trailing edge. However, some differences can be observed, which include a vortex merging process for the lowest Reynolds number, $Re = 0.5 \times 10^5$, and the fact that the main tone occurs at a unique frequency for almost the entire period analyzed. On the other hand, the more prominent tonal peaks for $Re = 1 \times 10^5$ alternate between the frequencies of secondary tones and vortex merging is not observed. Moreover, the instantaneous main frequency shifts to higher values compared to the lowest Reynolds number setup. The same feature is also observed in experiments \cite{Probsting2015_regimes}.

When the Reynolds number increases, flow transition leads to permanent bursting of coherent structures on the airfoil suction side. As a consequence, flow interaction among both sides of the airfoil is reduced and the pressure side bubble becomes permanent rather than intermittent. Then, this bubble near the trailing edge is responsible for the shedding of coherent structures and, consequently, for noise generation. For $Re = 2 \times 10^5$, which is located in a ``gray zone'' between different flow regimes as shown by \citet{Probsting2015_regimes}, the vortex shedding occurs rather erratically such that well-defined coherent structures are rarely observed. Thus, the noise generation process is not very efficient but it still occurs in the form of multiple tones. For the highest Reynolds number analyzed, $Re = 4 \times 10^5$, a well-behaved periodic vortex shedding is observed from the pressure side and noise is generated in the form of a single tonal peak.

\begin{acknowledgments}
	The authors acknowledge Fun\-da\-\c{c}\~{a}o de Amparo \`{a} Pesquisa do Estado de S\~{a}o Paulo, FAPESP, for supporting the present work under research Grants No.\ 2013/08293-7, 2018/11835-0 and 2021/06448-0.
	The authors also acknowledge ``Centro Nacional de Processamento de Alto Desempenho em São Paulo'' (CENAPAD-SP), Project No. 551, and the National Laboratory for Scientific Computing (LNCC/MCTI), Project SimTurb, for providing the HPC resources used in this work. Finally, Conselho Nacional de Desenvolvimento Científico e Tecnológico, CNPq, is also acknowledged for supporting this research under Grants No. 407842/2018-7 and 304335/2018-5.
\end{acknowledgments}

\bibliography{bibfile}

\begin{thebibliography}{43}%
\makeatletter
\providecommand \@ifxundefined [1]{%
 \@ifx{#1\undefined}
}%
\providecommand \@ifnum [1]{%
 \ifnum #1\expandafter \@firstoftwo
 \else \expandafter \@secondoftwo
 \fi
}%
\providecommand \@ifx [1]{%
 \ifx #1\expandafter \@firstoftwo
 \else \expandafter \@secondoftwo
 \fi
}%
\providecommand \natexlab [1]{#1}%
\providecommand \enquote  [1]{``#1''}%
\providecommand \bibnamefont  [1]{#1}%
\providecommand \bibfnamefont [1]{#1}%
\providecommand \citenamefont [1]{#1}%
\providecommand \href@noop [0]{\@secondoftwo}%
\providecommand \href [0]{\begingroup \@sanitize@url \@href}%
\providecommand \@href[1]{\@@startlink{#1}\@@href}%
\providecommand \@@href[1]{\endgroup#1\@@endlink}%
\providecommand \@sanitize@url [0]{\catcode `\\12\catcode `\$12\catcode
  `\&12\catcode `\#12\catcode `\^12\catcode `\_12\catcode `\%12\relax}%
\providecommand \@@startlink[1]{}%
\providecommand \@@endlink[0]{}%
\providecommand \url  [0]{\begingroup\@sanitize@url \@url }%
\providecommand \@url [1]{\endgroup\@href {#1}{\urlprefix }}%
\providecommand \urlprefix  [0]{URL }%
\providecommand \Eprint [0]{\href }%
\providecommand \doibase [0]{https://doi.org/}%
\providecommand \selectlanguage [0]{\@gobble}%
\providecommand \bibinfo  [0]{\@secondoftwo}%
\providecommand \bibfield  [0]{\@secondoftwo}%
\providecommand \translation [1]{[#1]}%
\providecommand \BibitemOpen [0]{}%
\providecommand \bibitemStop [0]{}%
\providecommand \bibitemNoStop [0]{.\EOS\space}%
\providecommand \EOS [0]{\spacefactor3000\relax}%
\providecommand \BibitemShut  [1]{\csname bibitem#1\endcsname}%
\let\auto@bib@innerbib\@empty
\bibitem [{\citenamefont {Paterson}\ \emph {et~al.}(1973)\citenamefont
  {Paterson}, \citenamefont {Vogt}, \citenamefont {Fink},\ and\ \citenamefont
  {Munch}}]{Paterson1973}%
  \BibitemOpen
  \bibfield  {author} {\bibinfo {author} {\bibfnamefont {R.}~\bibnamefont
  {Paterson}}, \bibinfo {author} {\bibfnamefont {P.}~\bibnamefont {Vogt}},
  \bibinfo {author} {\bibfnamefont {M.}~\bibnamefont {Fink}},\ and\ \bibinfo
  {author} {\bibfnamefont {C.}~\bibnamefont {Munch}},\ }\bibfield  {title}
  {\bibinfo {title} {Vortex noise of isolated airfoils},\ }\href@noop {}
  {\bibfield  {journal} {\bibinfo  {journal} {J. Aircraft}\ }\textbf {\bibinfo
  {volume} {10}},\ \bibinfo {pages} {296} (\bibinfo {year} {1973})}\BibitemShut
  {NoStop}%
\bibitem [{\citenamefont {Tam}(1974)}]{Tam1974}%
  \BibitemOpen
  \bibfield  {author} {\bibinfo {author} {\bibfnamefont {C.~K.~W.}\
  \bibnamefont {Tam}},\ }\bibfield  {title} {\bibinfo {title} {Discrete tones
  of isolated airfoils},\ }\href
  {https://doi.org/http://dx.doi.org/10.1121/1.1914682} {\bibfield  {journal}
  {\bibinfo  {journal} {J. Acoust. Soc. Am.}\ }\textbf {\bibinfo {volume}
  {55}},\ \bibinfo {pages} {1173} (\bibinfo {year} {1974})}\BibitemShut
  {NoStop}%
\bibitem [{\citenamefont {Arbey}\ and\ \citenamefont
  {Bataille}(1983)}]{Arbey1983}%
  \BibitemOpen
  \bibfield  {author} {\bibinfo {author} {\bibfnamefont {H.}~\bibnamefont
  {Arbey}}\ and\ \bibinfo {author} {\bibfnamefont {J.}~\bibnamefont
  {Bataille}},\ }\bibfield  {title} {\bibinfo {title} {Noise generated by
  airfoil profiles placed in a uniform laminar flow},\ }\href
  {https://doi.org/10.1017/S0022112083003201} {\bibfield  {journal} {\bibinfo
  {journal} {J. Fluid Mech.}\ }\textbf {\bibinfo {volume} {134}},\ \bibinfo
  {pages} {33} (\bibinfo {year} {1983})}\BibitemShut {NoStop}%
\bibitem [{\citenamefont {Lowson}\ \emph {et~al.}(1994)\citenamefont {Lowson},
  \citenamefont {Fiddes},\ and\ \citenamefont {Nash}}]{Lowson1994}%
  \BibitemOpen
  \bibfield  {author} {\bibinfo {author} {\bibfnamefont {M.~V.}\ \bibnamefont
  {Lowson}}, \bibinfo {author} {\bibfnamefont {S.~P.}\ \bibnamefont {Fiddes}},\
  and\ \bibinfo {author} {\bibfnamefont {E.~C.}\ \bibnamefont {Nash}},\
  }\bibfield  {title} {\bibinfo {title} {Laminar boundary layer aeroacoustic
  instabilities},\ }in\ \href@noop {} {\emph {\bibinfo {booktitle} {32nd AIAA
  Aerospace Sciences Meeting and Exhibit, AIAA-Paper 94-0358}}}\ (\bibinfo
  {year} {1994})\ pp.\ \bibinfo {pages} {1--10}\BibitemShut {NoStop}%
\bibitem [{\citenamefont {Nash}\ \emph {et~al.}(1999)\citenamefont {Nash},
  \citenamefont {Lowson},\ and\ \citenamefont {Mc{A}lpine}}]{Nash1999}%
  \BibitemOpen
  \bibfield  {author} {\bibinfo {author} {\bibfnamefont {E.}~\bibnamefont
  {Nash}}, \bibinfo {author} {\bibfnamefont {M.}~\bibnamefont {Lowson}},\ and\
  \bibinfo {author} {\bibfnamefont {A.}~\bibnamefont {Mc{A}lpine}},\ }\bibfield
   {title} {\bibinfo {title} {Boundary-layer instability noise on aerofoils},\
  }\href {https://doi.org/10.1017/S002211209800367X} {\bibfield  {journal}
  {\bibinfo  {journal} {J. Fluid Mech.}\ }\textbf {\bibinfo {volume} {382}},\
  \bibinfo {pages} {27} (\bibinfo {year} {1999})}\BibitemShut {NoStop}%
\bibitem [{\citenamefont {Desquesnes}\ \emph {et~al.}(2007)\citenamefont
  {Desquesnes}, \citenamefont {Terracol},\ and\ \citenamefont
  {Sagaut}}]{Desquesnes2007}%
  \BibitemOpen
  \bibfield  {author} {\bibinfo {author} {\bibfnamefont {G.}~\bibnamefont
  {Desquesnes}}, \bibinfo {author} {\bibfnamefont {M.}~\bibnamefont
  {Terracol}},\ and\ \bibinfo {author} {\bibfnamefont {P.}~\bibnamefont
  {Sagaut}},\ }\bibfield  {title} {\bibinfo {title} {Numerical investigation of
  the tone noise mechanism over laminar airfoils},\ }\href
  {https://doi.org/10.1017/S0022112007007896} {\bibfield  {journal} {\bibinfo
  {journal} {J. Fluid Mech.}\ }\textbf {\bibinfo {volume} {591}},\ \bibinfo
  {pages} {155} (\bibinfo {year} {2007})}\BibitemShut {NoStop}%
\bibitem [{\citenamefont {Jones}\ and\ \citenamefont
  {Sandberg}(2011)}]{Jones2011}%
  \BibitemOpen
  \bibfield  {author} {\bibinfo {author} {\bibfnamefont {L.~E.}\ \bibnamefont
  {Jones}}\ and\ \bibinfo {author} {\bibfnamefont {R.~D.}\ \bibnamefont
  {Sandberg}},\ }\bibfield  {title} {\bibinfo {title} {Numerical analysis of
  tonal airfoil self-noise and acoustic feedback-loops},\ }\href
  {https://doi.org/https://doi.org/10.1016/j.jsv.2011.07.009} {\bibfield
  {journal} {\bibinfo  {journal} {Journal of Sound and Vibration}\ }\textbf
  {\bibinfo {volume} {330}},\ \bibinfo {pages} {6137} (\bibinfo {year}
  {2011})}\BibitemShut {NoStop}%
\bibitem [{\citenamefont {{Fosas de Pando}}\ \emph {et~al.}(2014)\citenamefont
  {{Fosas de Pando}}, \citenamefont {Schmid},\ and\ \citenamefont
  {Sipp}}]{FosasdePando2014}%
  \BibitemOpen
  \bibfield  {author} {\bibinfo {author} {\bibfnamefont {M.}~\bibnamefont
  {{Fosas de Pando}}}, \bibinfo {author} {\bibfnamefont {P.~J.}\ \bibnamefont
  {Schmid}},\ and\ \bibinfo {author} {\bibfnamefont {D.}~\bibnamefont {Sipp}},\
  }\bibfield  {title} {\bibinfo {title} {A global analysis of tonal noise in
  flows around aerofoils},\ }\href {https://doi.org/10.1017/jfm.2014.356}
  {\bibfield  {journal} {\bibinfo  {journal} {J. Fluid Mech.}\ }\textbf
  {\bibinfo {volume} {754}},\ \bibinfo {pages} {5} (\bibinfo {year}
  {2014})}\BibitemShut {NoStop}%
\bibitem [{\citenamefont {Wu}\ \emph {et~al.}(2021)\citenamefont {Wu},
  \citenamefont {Sandberg},\ and\ \citenamefont {Moreau}}]{Wu2021}%
  \BibitemOpen
  \bibfield  {author} {\bibinfo {author} {\bibfnamefont {H.}~\bibnamefont
  {Wu}}, \bibinfo {author} {\bibfnamefont {R.~D.}\ \bibnamefont {Sandberg}},\
  and\ \bibinfo {author} {\bibfnamefont {S.}~\bibnamefont {Moreau}},\
  }\bibfield  {title} {\bibinfo {title} {Stability characteristics of different
  aerofoil flows at {R}ec=150,000 and the implications for aerofoil
  self-noise},\ }\href
  {https://doi.org/https://doi.org/10.1016/j.jsv.2021.116152} {\bibfield
  {journal} {\bibinfo  {journal} {Journal of Sound and Vibration}\ }\textbf
  {\bibinfo {volume} {506}},\ \bibinfo {pages} {116152} (\bibinfo {year}
  {2021})}\BibitemShut {NoStop}%
\bibitem [{\citenamefont {Ricciardi}\ \emph {et~al.}(2022)\citenamefont
  {Ricciardi}, \citenamefont {Wolf},\ and\ \citenamefont
  {Taira}}]{ricciardi2022JFM}%
  \BibitemOpen
  \bibfield  {author} {\bibinfo {author} {\bibfnamefont {T.~R.}\ \bibnamefont
  {Ricciardi}}, \bibinfo {author} {\bibfnamefont {W.~R.}\ \bibnamefont
  {Wolf}},\ and\ \bibinfo {author} {\bibfnamefont {K.}~\bibnamefont {Taira}},\
  }\bibfield  {title} {\bibinfo {title} {Transition, intermittency and phase
  interference effects in airfoil secondary tones and acoustic feedback loop},\
  }\href {https://doi.org/10.1017/jfm.2022.129} {\bibfield  {journal} {\bibinfo
   {journal} {Journal of Fluid Mechanics}\ }\textbf {\bibinfo {volume} {937}},\
  \bibinfo {pages} {A23} (\bibinfo {year} {2022})}\BibitemShut {NoStop}%
\bibitem [{\citenamefont {Golubev}\ \emph {et~al.}(2014)\citenamefont
  {Golubev}, \citenamefont {Nguyen}, \citenamefont {Mankbadi}, \citenamefont
  {Roger},\ and\ \citenamefont {Visbal}}]{Golubev2014}%
  \BibitemOpen
  \bibfield  {author} {\bibinfo {author} {\bibfnamefont {V.~V.}\ \bibnamefont
  {Golubev}}, \bibinfo {author} {\bibfnamefont {L.}~\bibnamefont {Nguyen}},
  \bibinfo {author} {\bibfnamefont {R.~R.}\ \bibnamefont {Mankbadi}}, \bibinfo
  {author} {\bibfnamefont {M.}~\bibnamefont {Roger}},\ and\ \bibinfo {author}
  {\bibfnamefont {M.~R.}\ \bibnamefont {Visbal}},\ }\bibfield  {title}
  {\bibinfo {title} {On flow-acoustic resonant interactions in transitional
  airfoils},\ }\href {https://doi.org/10.1260/1475-472X.13.1-2.1} {\bibfield
  {journal} {\bibinfo  {journal} {International Journal of Aeroacoustics}\
  }\textbf {\bibinfo {volume} {13}},\ \bibinfo {pages} {1} (\bibinfo {year}
  {2014})}\BibitemShut {NoStop}%
\bibitem [{\citenamefont {Sanjose}\ \emph {et~al.}(2019)\citenamefont
  {Sanjose}, \citenamefont {Towne}, \citenamefont {Jaiswal}, \citenamefont
  {Moreau}, \citenamefont {Lele},\ and\ \citenamefont {Mann}}]{Sanjose2018}%
  \BibitemOpen
  \bibfield  {author} {\bibinfo {author} {\bibfnamefont {M.}~\bibnamefont
  {Sanjose}}, \bibinfo {author} {\bibfnamefont {A.}~\bibnamefont {Towne}},
  \bibinfo {author} {\bibfnamefont {P.}~\bibnamefont {Jaiswal}}, \bibinfo
  {author} {\bibfnamefont {S.}~\bibnamefont {Moreau}}, \bibinfo {author}
  {\bibfnamefont {S.~K.}\ \bibnamefont {Lele}},\ and\ \bibinfo {author}
  {\bibfnamefont {A.}~\bibnamefont {Mann}},\ }\bibfield  {title} {\bibinfo
  {title} {Modal analysis of the laminar boundary layer instability and tonal
  noise of an airfoil at {R}eynolds number 150,000},\ }\href
  {https://doi.org/10.1177/1475472X18812798} {\bibfield  {journal} {\bibinfo
  {journal} {Int. J. Aeroacoustics}\ }\textbf {\bibinfo {volume} {18}},\
  \bibinfo {pages} {317} (\bibinfo {year} {2019})}\BibitemShut {NoStop}%
\bibitem [{\citenamefont {Gelot}\ and\ \citenamefont
  {Kim}(2020)}]{gelot_kim_2020}%
  \BibitemOpen
  \bibfield  {author} {\bibinfo {author} {\bibfnamefont {M.~B.~R.}\
  \bibnamefont {Gelot}}\ and\ \bibinfo {author} {\bibfnamefont {J.~W.}\
  \bibnamefont {Kim}},\ }\bibfield  {title} {\bibinfo {title} {Effect of
  serrated trailing edges on aerofoil tonal noise},\ }\href
  {https://doi.org/10.1017/jfm.2020.724} {\bibfield  {journal} {\bibinfo
  {journal} {Journal of Fluid Mechanics}\ }\textbf {\bibinfo {volume} {904}},\
  \bibinfo {pages} {A30} (\bibinfo {year} {2020})}\BibitemShut {NoStop}%
\bibitem [{\citenamefont {Nguyen}\ \emph {et~al.}(2021)\citenamefont {Nguyen},
  \citenamefont {Golubev}, \citenamefont {Mankbadi}, \citenamefont {Yakhina},\
  and\ \citenamefont {Roger}}]{Golubev2021}%
  \BibitemOpen
  \bibfield  {author} {\bibinfo {author} {\bibfnamefont {L.}~\bibnamefont
  {Nguyen}}, \bibinfo {author} {\bibfnamefont {V.}~\bibnamefont {Golubev}},
  \bibinfo {author} {\bibfnamefont {R.}~\bibnamefont {Mankbadi}}, \bibinfo
  {author} {\bibfnamefont {G.}~\bibnamefont {Yakhina}},\ and\ \bibinfo {author}
  {\bibfnamefont {M.}~\bibnamefont {Roger}},\ }\bibfield  {title} {\bibinfo
  {title} {Numerical investigation of tonal trailing-edge noise radiated by low
  {R}eynolds number airfoils},\ }\bibfield  {journal} {\bibinfo  {journal}
  {Applied Sciences}\ }\textbf {\bibinfo {volume} {11}},\ \href
  {https://doi.org/10.3390/app11052257} {10.3390/app11052257} (\bibinfo {year}
  {2021})\BibitemShut {NoStop}%
\bibitem [{\citenamefont {Plogmann}\ \emph {et~al.}(2013)\citenamefont
  {Plogmann}, \citenamefont {Herrig},\ and\ \citenamefont
  {W\"{u}rz}}]{Plogmann2013}%
  \BibitemOpen
  \bibfield  {author} {\bibinfo {author} {\bibfnamefont {B.}~\bibnamefont
  {Plogmann}}, \bibinfo {author} {\bibfnamefont {A.}~\bibnamefont {Herrig}},\
  and\ \bibinfo {author} {\bibfnamefont {W.}~\bibnamefont {W\"{u}rz}},\
  }\bibfield  {title} {{\selectlanguage {English}\bibinfo {title} {Experimental
  investigations of a trailing edge noise feedback mechanism on a {NACA}0012
  airfoil}},\ }\href {https://doi.org/10.1007/s00348-013-1480-z} {\bibfield
  {journal} {\bibinfo  {journal} {Exp. Fluids}\ }\textbf {\bibinfo {volume}
  {54}},\ \bibinfo {eid} {1480} (\bibinfo {year} {2013})}\BibitemShut {NoStop}%
\bibitem [{\citenamefont {Wolf}\ \emph
  {et~al.}(2012{\natexlab{a}})\citenamefont {Wolf}, \citenamefont {Azevedo},\
  and\ \citenamefont {Lele}}]{Wolf2012}%
  \BibitemOpen
  \bibfield  {author} {\bibinfo {author} {\bibfnamefont {W.~R.}\ \bibnamefont
  {Wolf}}, \bibinfo {author} {\bibfnamefont {J.~L.~F.}\ \bibnamefont
  {Azevedo}},\ and\ \bibinfo {author} {\bibfnamefont {S.~K.}\ \bibnamefont
  {Lele}},\ }\bibfield  {title} {\bibinfo {title} {Convective effects and the
  role of quadrupole sources for aerofoil aeroacoustics},\ }\href
  {https://doi.org/10.1017/jfm.2012.327} {\bibfield  {journal} {\bibinfo
  {journal} {J. Fluid Mech.}\ }\textbf {\bibinfo {volume} {708}},\ \bibinfo
  {pages} {502} (\bibinfo {year} {2012}{\natexlab{a}})}\BibitemShut {NoStop}%
\bibitem [{\citenamefont {Pr\"{o}bsting}\ and\ \citenamefont
  {Yarusevych}(2015)}]{Probsting2015_bubble}%
  \BibitemOpen
  \bibfield  {author} {\bibinfo {author} {\bibfnamefont {S.}~\bibnamefont
  {Pr\"{o}bsting}}\ and\ \bibinfo {author} {\bibfnamefont {S.}~\bibnamefont
  {Yarusevych}},\ }\bibfield  {title} {\bibinfo {title} {Laminar separation
  bubble development on an airfoil emitting tonal noise},\ }\href
  {https://doi.org/10.1017/jfm.2015.427} {\bibfield  {journal} {\bibinfo
  {journal} {J. Fluid Mech.}\ }\textbf {\bibinfo {volume} {780}},\ \bibinfo
  {pages} {167} (\bibinfo {year} {2015})}\BibitemShut {NoStop}%
\bibitem [{\citenamefont {Pr\"{o}bsting}\ \emph {et~al.}(2015)\citenamefont
  {Pr\"{o}bsting}, \citenamefont {Scarano},\ and\ \citenamefont
  {Morris}}]{Probsting2015_regimes}%
  \BibitemOpen
  \bibfield  {author} {\bibinfo {author} {\bibfnamefont {S.}~\bibnamefont
  {Pr\"{o}bsting}}, \bibinfo {author} {\bibfnamefont {F.}~\bibnamefont
  {Scarano}},\ and\ \bibinfo {author} {\bibfnamefont {S.~C.}\ \bibnamefont
  {Morris}},\ }\bibfield  {title} {\bibinfo {title} {Regimes of tonal noise on
  an airfoil at moderate {R}eynolds number},\ }\href
  {https://doi.org/10.1017/jfm.2015.475} {\bibfield  {journal} {\bibinfo
  {journal} {J. Fluid Mech.}\ }\textbf {\bibinfo {volume} {780}},\ \bibinfo
  {pages} {407} (\bibinfo {year} {2015})}\BibitemShut {NoStop}%
\bibitem [{\citenamefont {Yakhina}\ \emph {et~al.}(2020)\citenamefont
  {Yakhina}, \citenamefont {Roger}, \citenamefont {Moreau}, \citenamefont
  {Nguyen},\ and\ \citenamefont {Golubev}}]{Yakhina2020}%
  \BibitemOpen
  \bibfield  {author} {\bibinfo {author} {\bibfnamefont {G.}~\bibnamefont
  {Yakhina}}, \bibinfo {author} {\bibfnamefont {M.}~\bibnamefont {Roger}},
  \bibinfo {author} {\bibfnamefont {S.}~\bibnamefont {Moreau}}, \bibinfo
  {author} {\bibfnamefont {L.}~\bibnamefont {Nguyen}},\ and\ \bibinfo {author}
  {\bibfnamefont {V.}~\bibnamefont {Golubev}},\ }\bibfield  {title} {\bibinfo
  {title} {Experimental and analytical investigation of the tonal trailing-edge
  noise radiated by low {R}eynolds number aerofoils},\ }\href
  {https://doi.org/10.3390/acoustics2020018} {\bibfield  {journal} {\bibinfo
  {journal} {Acoustics}\ }\textbf {\bibinfo {volume} {2}},\ \bibinfo {pages}
  {293} (\bibinfo {year} {2020})}\BibitemShut {NoStop}%
\bibitem [{\citenamefont {Pr\"{o}bsting}\ \emph {et~al.}(2014)\citenamefont
  {Pr\"{o}bsting}, \citenamefont {Serpieri},\ and\ \citenamefont
  {Scarano}}]{Probsting2014}%
  \BibitemOpen
  \bibfield  {author} {\bibinfo {author} {\bibfnamefont {S.}~\bibnamefont
  {Pr\"{o}bsting}}, \bibinfo {author} {\bibfnamefont {J.}~\bibnamefont
  {Serpieri}},\ and\ \bibinfo {author} {\bibfnamefont {F.}~\bibnamefont
  {Scarano}},\ }\bibfield  {title} {\bibinfo {title} {Experimental
  investigation of aerofoil tonal noise generation},\ }\href
  {https://doi.org/10.1017/jfm.2014.156} {\bibfield  {journal} {\bibinfo
  {journal} {J. Fluid Mech.}\ }\textbf {\bibinfo {volume} {747}},\ \bibinfo
  {pages} {656} (\bibinfo {year} {2014})}\BibitemShut {NoStop}%
\bibitem [{\citenamefont {Ricciardi}\ \emph {et~al.}(2021)\citenamefont
  {Ricciardi}, \citenamefont {Wolf},\ and\ \citenamefont
  {Taira}}]{Ricciardi_Scitech2021}%
  \BibitemOpen
  \bibfield  {author} {\bibinfo {author} {\bibfnamefont {T.~R.}\ \bibnamefont
  {Ricciardi}}, \bibinfo {author} {\bibfnamefont {W.~R.}\ \bibnamefont
  {Wolf}},\ and\ \bibinfo {author} {\bibfnamefont {K.}~\bibnamefont {Taira}},\
  }\bibfield  {title} {\bibinfo {title} {Compressibility effects in airfoil
  secondary tones},\ }in\ \href {https://doi.org/10.2514/6.2021-0455} {\emph
  {\bibinfo {booktitle} {AIAA Scitech 2021 Forum, AIAA Paper 2021-0455}}}\
  (\bibinfo {year} {2021})\BibitemShut {NoStop}%
\bibitem [{\citenamefont {Kurelek}\ \emph {et~al.}(2019)\citenamefont
  {Kurelek}, \citenamefont {Yarusevych},\ and\ \citenamefont
  {Kotsonis}}]{yarusevich2019_merging}%
  \BibitemOpen
  \bibfield  {author} {\bibinfo {author} {\bibfnamefont {J.~W.}\ \bibnamefont
  {Kurelek}}, \bibinfo {author} {\bibfnamefont {S.}~\bibnamefont
  {Yarusevych}},\ and\ \bibinfo {author} {\bibfnamefont {M.}~\bibnamefont
  {Kotsonis}},\ }\bibfield  {title} {\bibinfo {title} {Vortex merging in a
  laminar separation bubble under natural and forced conditions},\ }\href@noop
  {} {\bibfield  {journal} {\bibinfo  {journal} {Phys. Rev. Fluids}\ }\textbf
  {\bibinfo {volume} {4}} (\bibinfo {year} {2019})}\BibitemShut {NoStop}%
\bibitem [{\citenamefont {Jones}\ \emph {et~al.}(2008)\citenamefont {Jones},
  \citenamefont {Sandberg},\ and\ \citenamefont {Sandham}}]{jones2008}%
  \BibitemOpen
  \bibfield  {author} {\bibinfo {author} {\bibfnamefont {L.~E.}\ \bibnamefont
  {Jones}}, \bibinfo {author} {\bibfnamefont {R.~D.}\ \bibnamefont
  {Sandberg}},\ and\ \bibinfo {author} {\bibfnamefont {N.~D.}\ \bibnamefont
  {Sandham}},\ }\bibfield  {title} {\bibinfo {title} {Direct numerical
  simulations of forced and unforced separation bubbles on an airfoil at
  incidence},\ }\href {https://doi.org/10.1017/S0022112008000864} {\bibfield
  {journal} {\bibinfo  {journal} {J. Fluid Mech.}\ }\textbf {\bibinfo {volume}
  {602}},\ \bibinfo {pages} {175} (\bibinfo {year} {2008})}\BibitemShut
  {NoStop}%
\bibitem [{\citenamefont {Jones}\ \emph {et~al.}(2010)\citenamefont {Jones},
  \citenamefont {Sandberg},\ and\ \citenamefont {Sandham}}]{jones2010}%
  \BibitemOpen
  \bibfield  {author} {\bibinfo {author} {\bibfnamefont {L.~E.}\ \bibnamefont
  {Jones}}, \bibinfo {author} {\bibfnamefont {R.~D.}\ \bibnamefont
  {Sandberg}},\ and\ \bibinfo {author} {\bibfnamefont {N.~D.}\ \bibnamefont
  {Sandham}},\ }\bibfield  {title} {\bibinfo {title} {Stability and receptivity
  characteristics of a laminar separation bubble on an aerofoil},\ }\href
  {https://doi.org/10.1017/S0022112009993089} {\bibfield  {journal} {\bibinfo
  {journal} {J. Fluid Mech.}\ }\textbf {\bibinfo {volume} {648}},\ \bibinfo
  {pages} {257} (\bibinfo {year} {2010})}\BibitemShut {NoStop}%
\bibitem [{\citenamefont {Kurelek}\ \emph {et~al.}(2016)\citenamefont
  {Kurelek}, \citenamefont {Lambert},\ and\ \citenamefont
  {Yarusevych}}]{yarusevich2016_coherent}%
  \BibitemOpen
  \bibfield  {author} {\bibinfo {author} {\bibfnamefont {J.~W.}\ \bibnamefont
  {Kurelek}}, \bibinfo {author} {\bibfnamefont {A.~R.}\ \bibnamefont
  {Lambert}},\ and\ \bibinfo {author} {\bibfnamefont {S.}~\bibnamefont
  {Yarusevych}},\ }\bibfield  {title} {\bibinfo {title} {Coherent structures in
  the transition process of a laminar separation bubble},\ }\href
  {https://doi.org/10.2514/1.J054820} {\bibfield  {journal} {\bibinfo
  {journal} {AIAA J.}\ }\textbf {\bibinfo {volume} {54}},\ \bibinfo {pages}
  {2295} (\bibinfo {year} {2016})}\BibitemShut {NoStop}%
\bibitem [{\citenamefont {Kurelek}\ \emph {et~al.}(2018)\citenamefont
  {Kurelek}, \citenamefont {Kotsonis},\ and\ \citenamefont
  {Yarusevych}}]{yarusevych2018_transition}%
  \BibitemOpen
  \bibfield  {author} {\bibinfo {author} {\bibfnamefont {J.~W.}\ \bibnamefont
  {Kurelek}}, \bibinfo {author} {\bibfnamefont {M.}~\bibnamefont {Kotsonis}},\
  and\ \bibinfo {author} {\bibfnamefont {S.}~\bibnamefont {Yarusevych}},\
  }\bibfield  {title} {\bibinfo {title} {Transition in a separation bubble
  under tonal and broadband acoustic excitation},\ }\href
  {https://doi.org/10.1017/jfm.2018.546} {\bibfield  {journal} {\bibinfo
  {journal} {J. Fluid Mech.}\ }\textbf {\bibinfo {volume} {853}},\ \bibinfo
  {pages} {1} (\bibinfo {year} {2018})}\BibitemShut {NoStop}%
\bibitem [{\citenamefont {Jaiswal}\ \emph {et~al.}(2022)\citenamefont
  {Jaiswal}, \citenamefont {Pasco}, \citenamefont {Yakhina},\ and\
  \citenamefont {Moreau}}]{jaiswal2022}%
  \BibitemOpen
  \bibfield  {author} {\bibinfo {author} {\bibfnamefont {P.}~\bibnamefont
  {Jaiswal}}, \bibinfo {author} {\bibfnamefont {Y.}~\bibnamefont {Pasco}},
  \bibinfo {author} {\bibfnamefont {G.}~\bibnamefont {Yakhina}},\ and\ \bibinfo
  {author} {\bibfnamefont {S.}~\bibnamefont {Moreau}},\ }\bibfield  {title}
  {\bibinfo {title} {Experimental investigation of aerofoil tonal noise at low
  {M}ach number},\ }\href {https://doi.org/10.1017/jfm.2021.1018} {\bibfield
  {journal} {\bibinfo  {journal} {Journal of Fluid Mechanics}\ }\textbf
  {\bibinfo {volume} {932}},\ \bibinfo {pages} {A37} (\bibinfo {year}
  {2022})}\BibitemShut {NoStop}%
\bibitem [{\citenamefont {Padois}\ \emph {et~al.}(2016)\citenamefont {Padois},
  \citenamefont {Laffay}, \citenamefont {Idier},\ and\ \citenamefont
  {Moreau}}]{Padois2016}%
  \BibitemOpen
  \bibfield  {author} {\bibinfo {author} {\bibfnamefont {T.}~\bibnamefont
  {Padois}}, \bibinfo {author} {\bibfnamefont {P.}~\bibnamefont {Laffay}},
  \bibinfo {author} {\bibfnamefont {A.}~\bibnamefont {Idier}},\ and\ \bibinfo
  {author} {\bibfnamefont {S.}~\bibnamefont {Moreau}},\ }\bibfield  {title}
  {\bibinfo {title} {Tonal noise of a controlled-diffusion airfoil at low angle
  of attack and {R}eynolds number},\ }\href {https://doi.org/10.1121/1.4958916}
  {\bibfield  {journal} {\bibinfo  {journal} {The Journal of the Acoustical
  Society of America}\ }\textbf {\bibinfo {volume} {140}},\ \bibinfo {pages}
  {EL113} (\bibinfo {year} {2016})}\BibitemShut {NoStop}%
\bibitem [{\citenamefont {Nagarajan}\ \emph {et~al.}(2003)\citenamefont
  {Nagarajan}, \citenamefont {Lele},\ and\ \citenamefont
  {Ferziger}}]{Nagarajan2003}%
  \BibitemOpen
  \bibfield  {author} {\bibinfo {author} {\bibfnamefont {S.}~\bibnamefont
  {Nagarajan}}, \bibinfo {author} {\bibfnamefont {S.~K.}\ \bibnamefont
  {Lele}},\ and\ \bibinfo {author} {\bibfnamefont {J.~H.}\ \bibnamefont
  {Ferziger}},\ }\bibfield  {title} {\bibinfo {title} {A robust high-order
  compact method for large eddy simulation},\ }\href
  {https://doi.org/10.1016/S0021-9991(03)00322-X} {\bibfield  {journal}
  {\bibinfo  {journal} {J. Comput. Phys.}\ }\textbf {\bibinfo {volume} {191}},\
  \bibinfo {pages} {392} (\bibinfo {year} {2003})}\BibitemShut {NoStop}%
\bibitem [{\citenamefont {Lele}(1992)}]{Lele1992}%
  \BibitemOpen
  \bibfield  {author} {\bibinfo {author} {\bibfnamefont {S.~K.}\ \bibnamefont
  {Lele}},\ }\bibfield  {title} {\bibinfo {title} {Compact finite difference
  schemes with spectral-like resolution},\ }\href
  {https://doi.org/http://dx.doi.org/10.1016/0021-9991(92)90324-R} {\bibfield
  {journal} {\bibinfo  {journal} {J. Comput. Phys.}\ }\textbf {\bibinfo
  {volume} {103}},\ \bibinfo {pages} {16} (\bibinfo {year} {1992})}\BibitemShut
  {NoStop}%
\bibitem [{\citenamefont {Mathew}\ \emph {et~al.}(2003)\citenamefont {Mathew},
  \citenamefont {Lechner}, \citenamefont {Foysi}, \citenamefont {Sesterhenn},\
  and\ \citenamefont {Friedrich}}]{Mathew_etal_2003}%
  \BibitemOpen
  \bibfield  {author} {\bibinfo {author} {\bibfnamefont {J.}~\bibnamefont
  {Mathew}}, \bibinfo {author} {\bibfnamefont {R.}~\bibnamefont {Lechner}},
  \bibinfo {author} {\bibfnamefont {H.}~\bibnamefont {Foysi}}, \bibinfo
  {author} {\bibfnamefont {J.}~\bibnamefont {Sesterhenn}},\ and\ \bibinfo
  {author} {\bibfnamefont {R.}~\bibnamefont {Friedrich}},\ }\bibfield  {title}
  {\bibinfo {title} {An explicit filtering method for large eddy simulation of
  compressible flows},\ }\href {https://doi.org/10.1063/1.1586271} {\bibfield
  {journal} {\bibinfo  {journal} {Physics of Fluids}\ }\textbf {\bibinfo
  {volume} {15}},\ \bibinfo {pages} {2279} (\bibinfo {year}
  {2003})}\BibitemShut {NoStop}%
\bibitem [{\citenamefont {Beam}\ and\ \citenamefont
  {Warming}(1978)}]{Beam1978}%
  \BibitemOpen
  \bibfield  {author} {\bibinfo {author} {\bibfnamefont {R.}~\bibnamefont
  {Beam}}\ and\ \bibinfo {author} {\bibfnamefont {R.}~\bibnamefont {Warming}},\
  }\bibfield  {title} {\bibinfo {title} {An implicit factored scheme for the
  compressible {N}avier-{S}tokes equations},\ }\href
  {https://doi.org/10.2514/3.60901} {\bibfield  {journal} {\bibinfo  {journal}
  {AIAA J.}\ }\textbf {\bibinfo {volume} {16}},\ \bibinfo {pages} {393}
  (\bibinfo {year} {1978})}\BibitemShut {NoStop}%
\bibitem [{\citenamefont {Bhaskaran}\ and\ \citenamefont
  {Lele}(2010)}]{Bhaskaran2010}%
  \BibitemOpen
  \bibfield  {author} {\bibinfo {author} {\bibfnamefont {R.}~\bibnamefont
  {Bhaskaran}}\ and\ \bibinfo {author} {\bibfnamefont {S.~K.}\ \bibnamefont
  {Lele}},\ }\bibfield  {title} {\bibinfo {title} {Large eddy simulation of
  free-stream turbulence effects on heat transfer to a high-pressure turbine
  cascade},\ }\href {https://doi.org/10.1080/14685241003705756} {\bibfield
  {journal} {\bibinfo  {journal} {J. Turb.}\ }\textbf {\bibinfo {volume}
  {11}},\ \bibinfo {pages} {N6} (\bibinfo {year} {2010})}\BibitemShut {NoStop}%
\bibitem [{\citenamefont {Wolf}\ \emph
  {et~al.}(2012{\natexlab{b}})\citenamefont {Wolf}, \citenamefont {Lele},
  \citenamefont {Jothiprasard},\ and\ \citenamefont {Cheung}}]{Wolf:DU96}%
  \BibitemOpen
  \bibfield  {author} {\bibinfo {author} {\bibfnamefont {W.~R.}\ \bibnamefont
  {Wolf}}, \bibinfo {author} {\bibfnamefont {S.~K.}\ \bibnamefont {Lele}},
  \bibinfo {author} {\bibfnamefont {G.}~\bibnamefont {Jothiprasard}},\ and\
  \bibinfo {author} {\bibfnamefont {L.}~\bibnamefont {Cheung}},\ }\bibfield
  {title} {\bibinfo {title} {Investigation of noise generated by a {DU}96
  airfoil},\ }in\ \href@noop {} {\emph {\bibinfo {booktitle} {18th AIAA/CEAS
  Aeroacoustics Conference (33th AIAA Aeroacoustics Conference), AIAA Paper
  2012-2055}}}\ (\bibinfo {year} {2012})\ pp.\ \bibinfo {pages}
  {1--15}\BibitemShut {NoStop}%
\bibitem [{\citenamefont {Wolf}\ \emph {et~al.}(2013)\citenamefont {Wolf},
  \citenamefont {Azevedo},\ and\ \citenamefont {Lele}}]{Wolf2013}%
  \BibitemOpen
  \bibfield  {author} {\bibinfo {author} {\bibfnamefont {W.~R.}\ \bibnamefont
  {Wolf}}, \bibinfo {author} {\bibfnamefont {J.~L.~F.}\ \bibnamefont
  {Azevedo}},\ and\ \bibinfo {author} {\bibfnamefont {S.~K.}\ \bibnamefont
  {Lele}},\ }\bibfield  {title} {\bibinfo {title} {Effects of mean flow
  convection, quadrupole sources and vortex shedding on airfoil overall sound
  pressure level},\ }\href {https://doi.org/10.1016/j.jsv.2013.08.029}
  {\bibfield  {journal} {\bibinfo  {journal} {J. Sound Vib.}\ }\textbf
  {\bibinfo {volume} {332}},\ \bibinfo {pages} {6905} (\bibinfo {year}
  {2013})}\BibitemShut {NoStop}%
\bibitem [{\citenamefont {Ramos}\ \emph {et~al.}(2019)\citenamefont {Ramos},
  \citenamefont {Wolf}, \citenamefont {Yeh},\ and\ \citenamefont
  {Taira}}]{Brener2019}%
  \BibitemOpen
  \bibfield  {author} {\bibinfo {author} {\bibfnamefont {B.~L.~O.}\
  \bibnamefont {Ramos}}, \bibinfo {author} {\bibfnamefont {W.~R.}\ \bibnamefont
  {Wolf}}, \bibinfo {author} {\bibfnamefont {C.-A.}\ \bibnamefont {Yeh}},\ and\
  \bibinfo {author} {\bibfnamefont {K.}~\bibnamefont {Taira}},\ }\bibfield
  {title} {\bibinfo {title} {Active flow control for drag reduction of a
  plunging airfoil under deep dynamic stall},\ }\href
  {https://doi.org/10.1103/PhysRevFluids.4.074603} {\bibfield  {journal}
  {\bibinfo  {journal} {Phys. Rev. Fluids}\ }\textbf {\bibinfo {volume} {4}},\
  \bibinfo {pages} {074603} (\bibinfo {year} {2019})}\BibitemShut {NoStop}%
\bibitem [{\citenamefont {Miotto}\ \emph {et~al.}(2022)\citenamefont {Miotto},
  \citenamefont {Wolf}, \citenamefont {Gaitonde},\ and\ \citenamefont
  {Visbal}}]{Miotto2022}%
  \BibitemOpen
  \bibfield  {author} {\bibinfo {author} {\bibfnamefont {R.~F.}\ \bibnamefont
  {Miotto}}, \bibinfo {author} {\bibfnamefont {W.~R.}\ \bibnamefont {Wolf}},
  \bibinfo {author} {\bibfnamefont {D.}~\bibnamefont {Gaitonde}},\ and\
  \bibinfo {author} {\bibfnamefont {M.}~\bibnamefont {Visbal}},\ }\bibfield
  {title} {\bibinfo {title} {Analysis of the onset and evolution of a dynamic
  stall vortex on a periodic plunging aerofoil},\ }\href
  {https://doi.org/10.1017/jfm.2022.165} {\bibfield  {journal} {\bibinfo
  {journal} {J. Fluid Mech.}\ }\textbf {\bibinfo {volume} {938}},\ \bibinfo
  {pages} {A24} (\bibinfo {year} {2022})}\BibitemShut {NoStop}%
\bibitem [{\citenamefont {Ricciardi}\ \emph {et~al.}(2020)\citenamefont
  {Ricciardi}, \citenamefont {Arias-Ramirez},\ and\ \citenamefont
  {Wolf}}]{Ricciardi2019_tones}%
  \BibitemOpen
  \bibfield  {author} {\bibinfo {author} {\bibfnamefont {T.~R.}\ \bibnamefont
  {Ricciardi}}, \bibinfo {author} {\bibfnamefont {W.}~\bibnamefont
  {Arias-Ramirez}},\ and\ \bibinfo {author} {\bibfnamefont {W.~R.}\
  \bibnamefont {Wolf}},\ }\bibfield  {title} {\bibinfo {title} {On secondary
  tones arising in trailing-edge noise at moderate {R}eynolds numbers},\ }\href
  {https://doi.org/10.1016/j.euromechflu.2019.08.015} {\bibfield  {journal}
  {\bibinfo  {journal} {Eur. J. Mech. B}\ }\textbf {\bibinfo {volume} {79}},\
  \bibinfo {pages} {54} (\bibinfo {year} {2020})}\BibitemShut {NoStop}%
\bibitem [{\citenamefont {Georgiadis}\ \emph {et~al.}(2010)\citenamefont
  {Georgiadis}, \citenamefont {Rizzetta},\ and\ \citenamefont
  {Fureby}}]{Georgiadis2010}%
  \BibitemOpen
  \bibfield  {author} {\bibinfo {author} {\bibfnamefont {N.~J.}\ \bibnamefont
  {Georgiadis}}, \bibinfo {author} {\bibfnamefont {D.~P.}\ \bibnamefont
  {Rizzetta}},\ and\ \bibinfo {author} {\bibfnamefont {C.}~\bibnamefont
  {Fureby}},\ }\bibfield  {title} {\bibinfo {title} {Large-eddy simulation:
  Current capabilities, recommended practices, and future research},\ }\href
  {https://doi.org/10.2514/1.J050232} {\bibfield  {journal} {\bibinfo
  {journal} {AIAA Journal}\ }\textbf {\bibinfo {volume} {48}},\ \bibinfo
  {pages} {1772} (\bibinfo {year} {2010})}\BibitemShut {NoStop}%
\bibitem [{\citenamefont {{Ffowcs Williams}}\ and\ \citenamefont
  {Hall}(1970)}]{FWHall1970}%
  \BibitemOpen
  \bibfield  {author} {\bibinfo {author} {\bibfnamefont {J.~E.}\ \bibnamefont
  {{Ffowcs Williams}}}\ and\ \bibinfo {author} {\bibfnamefont {L.~H.}\
  \bibnamefont {Hall}},\ }\bibfield  {title} {\bibinfo {title} {Aerodynamic
  sound generation by turbulent flow in the vicinity of a scattering half
  plane},\ }\href@noop {} {\bibfield  {journal} {\bibinfo  {journal} {Journal
  of Fluid Mechanics}\ }\textbf {\bibinfo {volume} {40}},\ \bibinfo {pages}
  {657} (\bibinfo {year} {1970})}\BibitemShut {NoStop}%
\bibitem [{\citenamefont {Sano}\ \emph {et~al.}(2019)\citenamefont {Sano},
  \citenamefont {Abreu}, \citenamefont {Cavalieri},\ and\ \citenamefont
  {Wolf}}]{sanoprf2019}%
  \BibitemOpen
  \bibfield  {author} {\bibinfo {author} {\bibfnamefont {A.}~\bibnamefont
  {Sano}}, \bibinfo {author} {\bibfnamefont {L.~I.}\ \bibnamefont {Abreu}},
  \bibinfo {author} {\bibfnamefont {A.~V.~G.}\ \bibnamefont {Cavalieri}},\ and\
  \bibinfo {author} {\bibfnamefont {W.~R.}\ \bibnamefont {Wolf}},\ }\bibfield
  {title} {\bibinfo {title} {Trailing-edge noise from the scattering of
  spanwise-coherent structures},\ }\href
  {https://doi.org/10.1103/PhysRevFluids.4.094602} {\bibfield  {journal}
  {\bibinfo  {journal} {Phys. Rev. Fluids}\ }\textbf {\bibinfo {volume} {4}},\
  \bibinfo {pages} {094602} (\bibinfo {year} {2019})}\BibitemShut {NoStop}%
\bibitem [{\citenamefont {Lumley}(1967)}]{Lumley1967}%
  \BibitemOpen
  \bibfield  {author} {\bibinfo {author} {\bibfnamefont {J.~L.}\ \bibnamefont
  {Lumley}},\ }\bibfield  {title} {\bibinfo {title} {The structure of
  inhomogeneous turbulent flows},\ }\href@noop {} {\bibfield  {journal}
  {\bibinfo  {journal} {Atmos. Turbul. Radio Wave Propag.}\ }\textbf {\bibinfo
  {volume} {790}},\ \bibinfo {pages} {166} (\bibinfo {year}
  {1967})}\BibitemShut {NoStop}%
\bibitem [{\citenamefont {Sirovich}(1987)}]{Sirovich1990}%
  \BibitemOpen
  \bibfield  {author} {\bibinfo {author} {\bibfnamefont {L.}~\bibnamefont
  {Sirovich}},\ }\bibfield  {title} {\bibinfo {title} {Turbulence and the
  dynamics of coherent structures. part {I}: Coherent structures},\ }\href@noop
  {} {\bibfield  {journal} {\bibinfo  {journal} {Quarterly of Applied
  Mathematics}\ }\textbf {\bibinfo {volume} {45}},\ \bibinfo {pages} {561}
  (\bibinfo {year} {1987})}\BibitemShut {NoStop}%
\end{thebibliography}%

\end{document}